\documentclass[12pt]{iopart}
\usepackage{amssymb}
\usepackage{diagbox} 
\usepackage{dcolumn}
\expandafter\let\csname equation*\endcsname\relax 
\expandafter\let\csname endequation*\endcsname\relax 
\usepackage{amsmath}
\usepackage[english]{babel}
\usepackage[latin1]{inputenc}
\usepackage[usenames, dvipsnames]{color}
\usepackage{graphicx}
\usepackage{gensymb}
\usepackage{iopams}
\usepackage{bm}        
\usepackage{color}
\usepackage{mathtools}
\usepackage{caption}
\usepackage{subcaption}
\usepackage{xcolor}
\usepackage{multirow,booktabs}
\usepackage{tabularx}

\newcolumntype{C}[1]{>{\centering\let\newline\\\arraybackslash\hspace{0pt}}m{#1}}

\begin{document}

\title{Multipartite entanglement in spin chains  and the hyperdeterminant}
\author{Alba Cervera-Lierta$^{1,2}$, Albert Gasull$^2$, Jos\'e I Latorre$^{2,3}$ and Germ\'an Sierra$^4$}
\address{$^1$ Barcelona Supercomputing Center (BSC).\\
$^2$ Dept. F\'isica Qu\`antica i Astrof\'isica, Universitat de Barcelona\\
$^3$ Center for Quantum Technologies, National University of Singapore.\\
$^4$ Instituto  de  F\'isica  Te\'orica  UAM/CSIC, Universidad  Aut\'onoma  de  Madrid,  Cantoblanco,  Madrid,  Spain.}
\ead{a.cervera.lierta@gmail.com}

\vspace{10pt}
\begin{indented}
\item[]\today
\end{indented}

\begin{abstract}
A way to characterize multipartite entanglement in  pure  states 
of a spin chain with $n$ sites and  
local dimension $d$ is by means of  the Cayley hyperdeterminant. 
The latter quantity   is a  polynomial  constructed with the components of the wave function $\psi_{i_1, \dots, i_n}$ which is invariant under local unitary transformation.
For spin 1/2 chains (i.e. $d=2$)  with $n=2$ and $n=3$ sites,  the hyperdeterminant coincides with  the 
concurrence and the tangle respectively.  In this paper we consider spin chains with $n=4$ sites
where  the hyperdeterminant is  a polynomial of degree 24 containing  around  $2.8 \times 10^6$ terms.  
This huge object  can be written in terms of more simple polynomials $S$ and $T$ of degrees 8 and 12 respectively.
Correspondingly we compute $S$, $T$ and the hyperdeterminant for eigenstates of the following 
spin chain Hamiltonians:  the transverse Ising model, the XXZ Heisenberg model and the Haldane-Shastry model. 
Those invariants are also  computed for random states, the ground states of random matrix Hamiltonians in the  Wigner-Dyson 
Gaussian ensembles and the  quadripartite entangled states defined by Verstraete et al. in 2002. 
Finally, we propose a generalization of the hyperdeterminant to thermal  density matrices.We observe how these polynomials are able to capture the phase transitions present in the models studied as well as a subclass of quadripartite entanglement present in the eigenstates. \\


\end{abstract}

\maketitle

\medmuskip=3mu
\thinmuskip=3mu
\thickmuskip=3mu

\section{Introduction}

Entanglement has been extensively studied in the context of condensed matter quantum systems \cite{qe}. It has proven useful to provide a deeper understanding of quantum phase transitions, as well as to validate the faithfulness of numerical approximations such as tensor networks \cite{cv}. 

 Most of the studies of entanglement are related to correlations among bi-partitions of a system. As a relevant example, we may consider the quantum correlations between two separate parts of a quantum system on a lattice using entanglement entropy as a figure of merit. 
 It has been found that most systems of interest obey the so called area law for the scaling of the entanglement entropy as the size of the part increases \cite{B86,S93,A08,E10,H07}.

We shall here focus on the study of entanglement in spin-$\frac{1}{2}$ chains. These one-dimensional systems present quantum phase transitions. The characterization of such critical behavior is determined by conformal symmetry. Indeed, at quantum phase transitions the system displays conformal invariance, and its analytic structure provides very powerful instruments to characterize correlations. Let us illustrate the power of conformal symmetry by considering the entanglement entropy corresponding to the reduced density matrix of a block of size $L$ out of $N$ ,
\begin{eqnarray}
	S(\rho_L)=-\mathrm{Tr}(\rho_L\log\rho_L),
\end{eqnarray} 
where $\rho_{L}=\mathrm{Tr}_{N-L}|\Psi_{g}\rangle\langle\Psi_{g}|$
and $|\Psi_{g}\rangle$ is the ground state of the system.
Then, it can be proven that this entanglement entropy
scales at a quantum phase transition as \cite{C94,H94,V03,C04}
\begin{eqnarray}
 S_{L}\sim \frac{c}{3}\log L, 
\end{eqnarray}
where $c$ is the central charge that defines the universality class of the model. Away from criticality, this entropy saturates to a constant 
that depends on the correlation length present in the system.

Many other different figures of merit for entanglement can be applied to spin chains. Nevertheless, some of them do not show scaling properties or fail to grab the subtleties of phase transitions. Entanglement entropy is a representative of figures of merit such as Renyi entropies, all of them obeying  scaling properties related to the universality class of the system. 

It is reasonable to look for a complete characterization of quantum correlations beyond the one provided by entanglement entropies. It is often argued that there is a need for new measures of {\sl genuine multipartite entanglement}. There is some ambiguity in the literature about this term. It is often referred as multipartite entanglement the study of correlations between two parties of a large system of particles \cite{Facchi,Facchi2,Uzo} . On the other hand, genuine multipartite entanglement can be referred as anything which analyzes correlations beyond two parties. There is a second more stringent definition that states that measures of genuine multipartite entanglement should not involve any partial trace of the system. This definition makes it very hard if not impossible to conduct studies in large systems. An example of a measure of strict multipartite entanglement could be the study of  Bell inequalities involving every party in a system. 

There are studies of multipartite entanglement in spin chains that involve figures of merit for three spins \cite{bayat}. This can be done using the {\sl tangle}, which corresponds to a hyperdeterminant of a tensor of three two-valued indices. Let us introduce a construction of the tangle as follows. Consider a quantum state made out of three qubits (spin-$\frac{1}{2}$)
\begin{eqnarray}
  |\psi\rangle = \sum_{i,j,k=0,1} b_{ijk} | i,j,k\rangle,
\end{eqnarray}
where the coefficients of the tensor fulfill a normalization condition $\sum_{i,j,k=0,1}b^*_{ijk}b_{ijk}=1$. The tangle of the state corresponds to the following polynomial of rank 4 \cite{Wootters}
\begin{eqnarray}
  \tau = 2 
  \vert b_{i_1 j_1 k_1}
  b_{i_2 j_2 k_2}b_{i_3 j_3 k_3}b_{i_4 j_4 k_4}
  \epsilon^{i_1 i_2}\epsilon^{j_1 j_2}\epsilon^{i_3 i_4}\epsilon^{j_3 j_4}
  \epsilon^{k_1 k_3} \epsilon^{k_2 k_4} \vert,
  \label{eq:tangle}
\end{eqnarray}
where all indices are contracted and $\epsilon^{ij}$ corresponds to the Levi-Civita tensor, i.e. $\epsilon^{00}=\epsilon^{11}=0$ and 
$\epsilon^{01}=-\epsilon^{10}=1$. Note that this contraction introduces minus signs, as opposed to pure contractions of subsystems which only involve the always positive Kronecker delta. The tangle is invariant under local unitary transformations on any party. It is a figure of genuine multipartite entanglement that involves no partition of the system. There are other works that study the multipartite entanglement in spin chains for an arbitrary, but finite, number of particles using the Meyer-Wallach measure of global entanglement \cite{Global}.

The purpose of this paper is to present a study of a figure of merit of multipartite entanglement based on the hyperdeterminant for 4 spins. The hyperdeterminant is a mathematical construction introduced by Cayley in the XIX century that serves the purpose of describing multipartite entanglement. The complexity to compute hyperdeterminants is remarkable and makes it difficult to apply it systematically to the study of quantum systems. Here, we shall introduce the basic properties of hyperdeterminants, its analysis for some special states and its behavior at a phase transition. 
 \cite{Nature,LatRicoVid,LatRiera,Nielsen,TesiDani,MEHdet}.
The content of the paper is organized as follows. In section \ref{sec:Hdet} we introduce the definition of hyperdeterminant and the so-called $S$ and $T$ invariants, as well as their generic interesting properties. Next, we extend these figures of merit to larger spin chains and for finite temperatures. Then, we study some interesting spin chain models such as the transverse Ising model  in section~\ref{sec:Ising}, the Heisenberg XXZ model  in section \ref{sec:XXZ}, and the  Haldane-Shastry model  in section \ref{sec:HS}.

\section{The Hyperdeterminant \label{sec:Hdet}}

The hyperdeterminant of a quantum pure state corresponds   
to a figure of merit for multipartite entanglement constructed as a polynomial of its coefficients. Given a pure state
\begin{eqnarray}
  |\psi\rangle = \sum_{i_1,\ldots,i_n}t_{i_1\ldots i_n}|i_1,\ldots,i_n\rangle, 
\end{eqnarray}
where the coefficients $t_{i_1\ldots i_n}$ form a complex tensor of $n$ indices
that  obey the normalization condition $\sum_{i_1\ldots i_n} t^*_ {i_1\ldots i_n} t_{i_1\ldots i_n}=1$.
The $n$-hyperdeterminant will be denoted as $\text{HDet}_{\text{n}}(t)$.
For $n=3$ spins, the hyperdeterminant ${\rm HDet}_3$ corresponds to the {\sl tangle} (often
called Wootters' tangle or three-tangle) \cite{Wootters}. Here we shall be interested in the case of four spins, that is on the study of ${\rm HDet}_4$.

\subsection{Definition and construction of the hyperdeterminant}

The hyperdeterminant was first introduced by A. Cayley \cite{Cayley} to characterize the condition for a system of linear equations
to have a non-trivial solution. To be precise, let us consider the case of a system of four equations
\begin{eqnarray}
 t_{ijkl} u_i v_j w_k & = & 0\nonumber\\
 t_{ijkl} u_i v_j z_l &= & 0\nonumber\\
 t_{ijkl} u_i w_k z_l & = & 0\nonumber\\
 t_{ijkl} v_j w_k z_l & = & 0,
  \label{eq:defhdet}
\end{eqnarray}
where all indices are contracted, $t$ is a tensor of four indices $i,j,k,l$ that run from 0 to 1, and $u, v, w$ and $z$ are two component vectors. As with the tangle definition of Eq.\eqref{eq:tangle}, all indices are contracted. 
The condition for the  above system of equations to  have a nontrivial solution is characterized by the hyperdeterminant
\begin{eqnarray}
  {\rm Non-trivial}\quad u,v,w,z \qquad  {\rm iff}\qquad  {\rm HDet}_4(t) =0.
\end{eqnarray}
The hyperdeterminant generalizes the familiar concept of a determinant for tensors with only two indices. The above definition brings the intuition that the hyperdeterminant must be a homogeneous polynomial in the coefficients of the tensor.

The above definition of hyperdeterminant is valid for tensors of any number of indices, but it does not provide its explicit construction. Cayley found a generating formula for the rank of the polynomial as a function of the local dimension of each index and the number of indices (see table \ref{Tab:Hdet} for some examples).
In the case of four indices,  ${\rm HDet}_4$ is a polynomial of degree 24 with 2 894 276 terms. 

\begin{table}[h!]
\caption{\label{Tab:Hdet}Degrees and numbers of terms of the ${\rm HDet}_n$ as a function of the numbers of indices $n$ and their local dimension $d$. The case studied in this paper is shown in italic face.}
\begin{indented}
\item[]
\begin{tabular}{@{}llll}
\br
$\mathbf{n}$ & $\mathbf{d}$ & \textbf{Degree} & \textbf{\# terms} \\
\mr
2 & 2 & 2 & 2 \\
3 & 2 & 4 & 21 \\
{\it 4} & {\it 2} & {\it 24} & {\it 2 894 276} \\
2 & 3 & 3 & 6 \\
3 & 3 & 36 & unknown \\
3 & 4 & 1236 & unknown \\ 
\br
\end{tabular}
\end{indented}
\end{table}

Explicit expressions for hyperdeterminants are hard to obtain. Cayley gave the first expression for ${\rm HDet}_3$ \cite{Cayley}. 
Later on, Schl\"afli made the extension to ${\rm HDet}_4$ \cite{Schlafli}. The hyperdeterminant was first proposed as a measure of 4-qubit entanglement in \cite{MEHdet}. It is also related with other polynomial invariants \cite{PolyHdet} used too to quantify quadripartite entanglement. However, it fails to detect the entanglement present in, for example, the GHZ-type states, and so it can not be considered a genuine measure for all kinds of quadripartite entanglement.

There are several methods to compute the hyperdeterminant that can be found in \cite{Schlafli,PolyHdet,TesiDani}. The Schl\"afli original method consists on the computation of polynomials obtained from determinants of hypermatrices so that their discriminants correspond to the concurrence, three-tangle and hyperdeterminant, depending on the dimensions of the tensor under discussion.

Let's start with a generic $2\times 2$ matrix $C$. Its hyperdeterminant, $\mathrm{HDet}_{2}$, corresponds to its determinant, $c_{00}c_{11}-c_{10}c_{01}$. If we identify each matrix element as the coefficients of a two qubits wave function, i.e. for some two qubits state $|\psi\rangle=\sum_{i,j=0,1}c_{ij}|ij\rangle$, $\mathrm{HDet}_{2}$ corresponds to the concurrence of this state. The next step is obtaining $\mathrm{HDet}_{3}$ by replacing each $c_{ij}$ coefficient with $b_{ij0}+b_{ij1} x$ in the $\mathrm{HDet}_{2}$ expression and computing the discriminant of the polynomial obtained, namely $P_{3}(x)$. If we identify each $b_{ijk}$ element with the coefficient of a three qubits state, $|\phi\rangle=\sum_{i,j,k=0,1}b_{ijk}|ijk\rangle$, then $\mathrm{HDet}_{3}$ corresponds to the tangle. Finally, we continue with this process one more time to obtain the hyperdeterminant of degree 4, $\mathrm{HDet}_{4}$. First replacing the $b_{ijk}$ coefficients of the previous $\mathrm{HDet}_{3}$ expression by $t_{ijk0}+t_{ijk1} x$ and second computing the discriminant of the polynomial obtained of degree four $P_{4}(x)$. Coefficients $t_{ijkl}$ are the same as the ones used in Eq.\eqref{eq:defhdet} and we can identify them with the wave function coefficients of a four qubits state, i.e. $|\varphi\rangle=\sum_{i,j,k,l=0,1}t_{ijkl}|ijkl\rangle$.

A discriminant could be complex or real, depending on the coefficients of the polynomial. If the coefficients are real numbers, then the discriminant is always real. In that case, it is zero if at least two roots are equal; it is positive if there exist $2k$ pairs of conjugate roots for $0\leq k\leq n/2$ where $n$ is the degree of the polynomial; and it is negative if there exist $2k+1$ pairs of conjugate roots for $0\leq k\leq (n-2)/4$ \cite{discriminants}. Then, it is possible to obtain a concurrence, tangle or $\mathrm{HDet}_{4}$ complex: if this is the case, we will take the absolute value, as it is done in previous works with the tangle \cite{Wootters}.


For 4-qubit system we can also compute the hyperdeterminant from the two polynomial invariants $S$ and $T$ \cite{PolyHdet}. Once we have obtained the polynomial of degree four $P_4(x)$, we identify the coefficients of this polynomial, that is $P_{4}(x)= b_0x^4+4b_{1}x^3+6b_{2}x^2+4b_{3}x+b_4$. Then, these invariants take the form
\begin{eqnarray}
S&=&3b_{2}^2-4b_{1}b_{3}+b_{0}b_{4}, \\
T&=&-b_{2}^3+2b_{1}b_{2}b_{3}-b_{0}b_{3}^2-b_{1}^2b_{4}+b_{0}b_{2}b_{4}.
\end{eqnarray}
Then, the hyperdeterminant is obtained as the following combination
\begin{eqnarray}
\mathrm{HDet}_{4}(|\Psi\rangle)=S^3-27 T^2.
\label{eq:ST}
\end{eqnarray}

There is a connection between the hyperdeterminant and the theory of elliptic curves \cite{Gibbs}: the $J$-invariant of an elliptic curve, an independent quantity which is invariant under rational transformations, can be expressed as $J=S^3/\mathrm{HDet}_{4}$. There is a known connection between hyperdeterminants and string theory: see for instance \cite{Duff,Duff2}.

\subsection{Basic properties of the hyperdeterminant}

The very definition of hyperdeterminant in Eq.\eqref{eq:defhdet} indicates that the hyperdeterminant is invariant under local changes of basis. That is, given a state $|\varphi\rangle$ and a state $|\tilde{\varphi}\rangle=U_1\otimes\cdots\otimes U_n|\varphi\rangle$, where $U_i$ are independent unitary changes of each local basis,
\begin{eqnarray}
{\rm HDet}_{\text{n}}(|\varphi\rangle)={\rm HDet}_{\text{n}}(|\tilde{\varphi}\rangle).
\end{eqnarray}
This immediately shows that
the hyperdeterminant provides a possible figure or merit to quantify  multipartite entanglement. 

The natural growth of complexity  in the study of 4-party entanglement is illustrated by the existence of 
 9 SLOCC classes of  pure 4-qubit states \cite{Verstraete}. Then, it is not surprising the existence of multiple non-equivalent figures of merit to quantify  multipartite entanglement. In \cite{PolyHdet},  the polynomial invariants of these 9 classes of 4-qubit states are computed and related to the hyperdeterminant, and the  $S$ and $T$ invariants.

It is worth remarking that the hyperdeterminant vanishes for  quantum states
that can be written as the product states on any bipartition. 
That is, for a state made out of four parties,
\begin{eqnarray}
|\psi\rangle &=&|\varphi\rangle_{1}|\phi\rangle_{234} \Rightarrow \mathrm{HDet}_{4}(|\psi\rangle)=0,\label{eq:1_234}\\
|\psi\rangle &=&|\varphi\rangle_{12}|\phi\rangle_{34} \ \Rightarrow \mathrm{HDet}_{4}(|\psi\rangle)=0,
\label{eq:12_34}
\end{eqnarray}
with the same result for any permutation of indices.
In the first case, when the state is a product state of 1-qubit and a generic state of the rest, the invariants $S$ and $T$ are zero, so is the hyperdeterminant. This brings the idea that the hyperdeterminant is only sensitive to genuine 4-party entanglement. In the second case, where the state can be separable in two halves,   
some more basic polynomial invariants are proportional to the concurrence, but it remains true that  $S$ and $T$  are zero, as well as the hyperdeterminant. 

\subsection{Definition of hyperdeterminant for mixed states}

We define the hyperdeterminant for a density matrix as follows. A density matrix can be expressed in its diagonal form as $\rho=\sum_{i}\lambda_{i}|\varphi_{i}\rangle\langle\varphi_{i}|$, where $\lambda_{i}$ are the eigenvalues and $|\varphi_{i}\rangle$ the eigenvectors of the matrix. 
%
Given all possible decompositions of $\rho$,
\begin{eqnarray}
\mathrm{HDet}_{4}(\rho)\equiv \min\sum_{i}\lambda_{i}\mathrm{HDet}_{4}(|\varphi_{i}\rangle,
\label{eq:HDetth2}
\end{eqnarray}
which an extension of the definition of Entanglement of Formation \cite{EntFor} for other figure of merit such as $\mathrm{HDet}_{4}$. We can extend the above definitions to $S$ and $T$ invariants.

The construction of hyperdeterminants for density matrices brings the possibility of defining the hyperdeterminant of  thermal states. 

Let us consider the density matrix of a system of $n$ spins  in equilibrium with a thermal reservoir
\begin{eqnarray}
\rho_{\beta}=\frac{\rme^{-\beta H}}{\mathcal{Z}}=\frac{1}{\mathcal{Z}}\sum_{i=0}^{2^n-1}\rme^{-\beta E_{i}}|E_{i}\rangle
\langle E_i| ,
\label{ther}
\end{eqnarray}
where $\mathcal{Z}=\mathrm{Tr}\left(\rme^{-\beta H}\right)$ is the partition function,  and $|E_{i}\rangle$ is the  state with energy $E_{i}$. We shall  define the hyperdeterminant of the thermal state (\ref{ther}) as 

\begin{eqnarray}
\mathrm{HDet}_4(\rho_{\beta})&\equiv&\frac{1}{\mathcal{Z}}\sum_{i=0}^{2^n-1}\rme^{-\beta E_{i}}\mathrm{HDet}_4(|E_{i}\rangle) 
\label{eq:Th}
\end{eqnarray}
where 
$\mathrm{HDet}_4(|E_{i}\rangle)$ is the hyperdeterminant of the state $|E_{i}\rangle$. 

In the case of degeneracy, a linear superposition of states with the same energy is also an eigenstate of the system. Then, the most general thermal state can be written as
\begin{equation}
|\psi\rangle_{th}=\frac{1}{\mathcal{Z}}\sum_{i}e^{-\beta E_{i}}\left(\sum_{j}a^{i}_{j}|E^{i}_{j}\rangle\right),
\end{equation}
where the first summation is over all different values of $E_{i}$ and the second corresponds to the linear superposition of eigenstates with the same eigenvalue $E_{i}$, with $\sum_{j}|a^{i}_{j}|^2=1$. Then, taking the second definition for $\mathrm{HDet}_4$ for mixed states \eqref{eq:HDetth2},
\begin{eqnarray}
\mathrm{HDet}_4(\rho_{\beta})&\equiv& \min_{a^{i}_{j}}\mathrm{HDet}_4(|\psi\rangle_{th}).
\label{eq:Th2}
\end{eqnarray}
A similar definitions hold for the thermal values of $S$ and $T$ invariants. 


\subsection{Examples}

\subsubsection{Special states}

We shall now  illustrate the computation of ${\rm HDet}_4$ for several special states. 

There are states for which the $\mathrm{HDet}_{4}$ vanishes because of the cancellation of $S$ and $T$ invariants. The most relevant example is the GHZ-like state \cite{GHZ}
\begin{eqnarray}
|GHZ\rangle=\frac{1}{\sqrt{2}}\left(|0000\rangle + |1111\rangle\right),
\label{ghz} 
\end{eqnarray}
which has $S=1/(2^6\cdot 3)$, $T=-1/(2^9\cdot 3^3)$ and zero $\mathrm{HDet}_{4}$. This result 
shows that $\mathrm{HDet}_{4}$ captures a different type of entanglement that the one associated to superposition of fully orthogonal states. 

There are other special states that have the same values as above for $S$ and $T$ invariants.
One example are the cluster states $|C_{1}\rangle$, $|C_{2}\rangle$ and $|C_{3}\rangle$ \cite{Cluster1,Cluster2},
\begin{eqnarray}
|C_{1}\rangle &=&\frac{1}{2}\left(|0000\rangle +|0011\rangle + |1100\rangle - |1111\rangle\right), 
\label{cluster} 
\\
|C_{2}\rangle &=&\frac{1}{2}\left(|0000\rangle +|0110\rangle + |1001\rangle - |1111\rangle\right), \\
|C_{3}\rangle &=&\frac{1}{2}\left(|0000\rangle +|0101\rangle + |1010\rangle - |1111\rangle\right), 
\end{eqnarray}
which maximizes the Von Neumann entropy of two of their three bipartition. Other example is the $|YC\rangle$ state \cite{YC},

\begin{eqnarray}
\fl |YC\rangle=\frac{1}{\sqrt{8}}\left(|0000\rangle -|0011\rangle -|0101\rangle + |0110\rangle 
+|1001\rangle + |1010\rangle + |1100\rangle +|1111\rangle \right),
\label{yc} 
\end{eqnarray}
which can perform a faithful teleportation of an arbitrary two-qubit entangled state. These states bring the idea that invariants $S$ and $T$ measure some kind of entanglement, but the hyperdeterminant makes a further selection. 

The $W$ state \cite{W},
\begin{eqnarray}
|W\rangle&=& \frac{1}{2}\left(|0001\rangle + |0010\rangle + |0100\rangle + |1000\rangle\right),
\label{eq:W}
\end{eqnarray}
has $S=T=0$. Again, W-ness is a different kind of entanglement as the one capture by $\mathrm{HDet}_4=0$.

Let us recall that 18 entanglement independent invariants are needed to classify four-qubit states under local unitaries \cite{PolyHdet}.  Most of these invariants are related to bi-partitions of the system, whereas $S$, $T$ and its combination into the $\mathrm{HDet}_{4}$ are measuring  global correlations involving every spin in the system.

On the other hand, states that maximize the $\mathrm{HDet}_{4}$ have been studied previously. Numerical analysis shows that a state with maximum $\mathrm{HDet}_{4}$ is \cite{Osterloh,Alsina}
\begin{eqnarray}
|HD\rangle=\frac{1}{\sqrt{6}}\left(|1000\rangle+|0100\rangle +|0010\rangle +|0001\rangle+\sqrt{2}|1111\rangle\right),
\label{hd} 
\end{eqnarray}
with $\mathrm{HDet}_4=1/(2^8\cdot 3^9)\simeq 1.98 \ 10^{-7}$, $S=0$ and $T=-1/(2^4 \cdot 3^6)$. Another state with the same values for the hyperdeterminant, $S$ and $T$ corresponds to the state $|L\rangle$ \cite{Cluster2}
\begin{eqnarray}
\fl |L\rangle &=&  \frac{1}{\sqrt{12}}\left[\left(1+w\right)\left(|0000\rangle+|1111\rangle\right) + \left(1-w\right)\left(|0011\rangle+|1100\rangle\right) \right. \nonumber\\ \fl  &+& \left. w^2\left(|0101\rangle+|0110\rangle+|1001\rangle+|1010\rangle\right)\right],
\end{eqnarray}
where $w=\exp(2\rmi\pi/3)$. This state also maximizes the average Tsallis entropy \cite{Tsallis} for $0<\alpha<2$ and $\alpha>2$. 

Other relevant states are the nine families of quadripartite entangled states defined by Verstraete {\sl et al.} in \cite{Verstraete}. The analysis of $\mathrm{HDet}_{4}$, $S$ and $T$ invariants for these families of states is  collected in \ref{app:HDet}.

\subsubsection{Random states}

In order to obtain a better picture of what are the typical values for $\mathrm{HDet}_{4}$, $S$ and $T$ invariants, we compute them for random pure states. The very definition of a random state
depends on the prior which is accepted. Here, we take as a prior two distributions of coefficients in the computational basis: a flat distribution, taking state coefficients with a random real and imaginary part and subject to the proper normalization of the state, and a Haar distribution, taking complex gaussian variables $z_{i}$, with zero median and unit variance.
Other options are perfectly valid, but we do not investigate them here.

We have generated 10000 random 4-qubit states with a flat and Haar prior on the coefficients and plotted $\mathrm{HDet}_{4}$ in figure \ref{fig:hdet-random} in comparison with ground state of random matrix  Hamiltonians that satisfy the GOE, GUE and GSE distributions
(see subsection {\ref{sec:GSrandom} ). 
The mean value of $\mathrm{HDet}_{4}$ is around $\sim 1.2\cdot 10^{-9}$, two orders of magnitude lower than the maximum possible value ($1.98\cdot 10^{-7}$ for $|HD\rangle$ state). Also, only $2\%$ of the states have $\mathrm{HDet}_{4}$ greater than $10^{-8}$. Similar results were obtained in \cite{TesiDani}. This result  is to be compared with the entanglement entropy of such states for a random bipartition, where maximal volume entropy is found. The $\mathrm{HDet}_{4}$ distribution obtained is not the same for flat and Haar distributed random states: the second have lower values of $\mathrm{HDet}_{4}$. Therefore, the hyperdeterminant is a more subtle figure of merit that is not maximal for most states, 
except for a small subset of random states, and can distinguish between two random priors.

A way to understand the scarce abundance of high hyperdeterminant states 
is based on the comparison between the 
multipartite  and the bipartite entanglements. 
The latter is 
measured by  the Von Neumann entropy, where one does not encounter cancellations coming from the 
different terms of the reduced density matrix. 
On the other hand, to obtain high hyperdeterminant values, requires a fine tuning to avoid 
cancellations. Random states do not propitiate those  cancellations that leads to 
low values for the hyperdeterminant. 


\subsubsection{Ground state of random Gaussian Hamiltonians }
\label{sec:GSrandom}


We construct a random matrix of dimension $2^n\times 2^n$ for $n=4$ whose entries are random numbers distributed following three types of Gaussian ensembles: Gaussian unitary ensemble (GUE), Gaussian orthogonal ensemble (GOE) and Gaussian symplectic ensemble (GSE).

Figure \ref{fig:hdet-random} shows the values of $\mathrm{HDet}_{4}$ for the ground state of $10^5$ random Hamiltonians for the three Gaussian distributions. For GUE and GSE, the mean value for $\mathrm{HDet}_{4}$ is slightly lower than for a random state and have the same value as Haar distributed random states, whereas for GOE is much smaller. This result is independent of the number of distributions considered, which suggests the existence of a probability density related to  $\mathrm{HDet}_{4}$.

\begin{figure}[t!]
\centering
\includegraphics[width=0.6\textwidth]{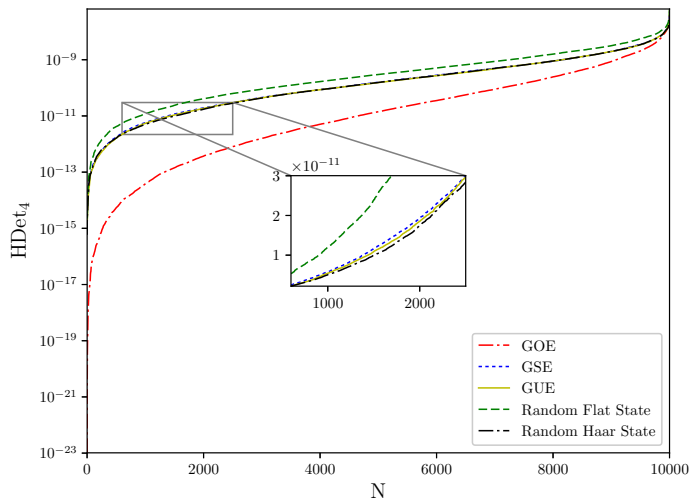}
\caption{$\mathrm{HDet}_{4}$ for $10^5$ random Hamiltonians distributed following a random distributions corresponding to the Gaussian unitary ensemble (GUE), the Gaussian orthogonal ensemble (GOE) and the Gaussian symplectic ensemble (GSE).These distributions are compared with $\mathrm{HDet}_{4}$ of flat and Haar distributed random states.}
\label{fig:hdet-random}
\end{figure}


\section{The transverse Ising model \label{sec:Ising}}

One of the most studied  1D  quantum spin chains is the transverse Ising model \cite{Review}. This model is described by the Hamiltonian
\begin{eqnarray}
H_{\text{Ising}}=-J\sum_{i=1}^n  \sigma_{i}^{x}\sigma_{i+1}^{x} - \lambda\sum_{i=1}^n \sigma_{i}^{z}.
\label{eq:HIsing}
\end{eqnarray}
where the sum is taken over the $n$ spins of a chain with periodic boundary conditions. 
We study the ferromagnetic interaction, i.e. $J>0$, and without lost of generality we set $J=1$ and $\lambda\geq 0$.

The non-commuting transverse field term introduces quantum fluctuations in the model causing a quantum phase transition from an ordered phase (magnetization different from zero) to a disordered paramagnetic phase (magnetization is zero), at critical value of $\lambda=\lambda_{c}$. 

For  infinite chains, $\lambda_{c}=1$ is the critical point where conformal invariance is restored. At $\lambda=0$ there are two degenerate ground states with ferromagnetic ordering, $|\rightarrow\rightarrow\cdots\rightarrow\rangle$ and $|\leftarrow\leftarrow\cdots\leftarrow\rangle$, where $|\rightarrow\rangle$ and $|\leftarrow\rangle$ are the spin states in the $\sigma^{x}$ basis, and at $\lambda>\lambda_{c}$ the external field strength wins over the neighboring interaction $J$ and the system lies in the paramagnetic phase. 

For  finite chains in the ferromagnetic phase, a non vanishing value of  $\lambda$ breaks the degeneracy of the ground state and produces an exponentially small energy gap between the two lowest energy states. On the other hand, the critical value 
$\lambda_c$ moves away from its value in the following sense. 
 The entropy of the  Ising spin chain peaks around the quantum phase transition. As long as the length of the chain increases, the critical point approaches to $1$. The entanglement entropy near $\lambda=1$ scales logarithmically following the conformal scaling law  with central charge $c=\frac{1}{2}$ till the correlation length bounds the entropy.



\subsection{Eigenstates}

We shall now compute the energy levels of the transverse Ising model for $n=4$ spins. The corresponding energies as a function of the external field $\lambda$ are
\begin{eqnarray}
\fl \left\lbrace -2\sqrt{2}\sqrt{\lambda'+\sqrt{\lambda''}}, \ -2\left(\sqrt{\lambda'}+1\right), \ -2\sqrt{2}\sqrt{\lambda'-\sqrt{\lambda''}}, \ -2\lambda, \ -2\lambda, -2\left(\sqrt{\lambda'}-1\right),  \right. \nonumber\\ \fl
 \left. 0, \ 0, \ 0, \ 0, \ 2\left(\sqrt{\lambda'}-1\right), \ 2\lambda, \ 2\lambda, \ 2\sqrt{2}\sqrt{\lambda'-\sqrt{\lambda''}}, 2\left(\sqrt{\lambda'}+1\right), \ 2\sqrt{2}\sqrt{\lambda'+\sqrt{\lambda''}} \right\rbrace.
\label{eq:Ising_energy}
\end{eqnarray}
where $\lambda'=1+\lambda^2$ and $\lambda''=1+\lambda^4$. The above eigenstates are ordered from the ground state to the $15^{th}$ excited state for $0<\lambda<2/\sqrt{3}$: for higher $\lambda$ some levels begin to cross each other, except the ground state and $15^{th}$ excited state, which remain the lowest and the highest energy levels respectively.

The analytic expressions of $\mathrm{HDet}_{4}$, $S$ and $T$ invariants for all the  eigenstates  
are summarized in table \ref{Tab:IsingTh} (see Appendix \ref{app:Ising} for details). 
One can distinguish three types of behaviors: 
{\sl i}) $\mathrm{HDet}_{4}$ is different from zero, {\sl ii}) $\mathrm{HDet}_{4}$ zero, due to a cancellation of non-vanishing $S$ and $T$ invariants, and {\sl iii}) $\mathrm{HDet}_{4}$, $S$ and $T$ are all zero.\\

\begin{table}
\begin{footnotesize}
\caption{\footnotesize\label{Tab:IsingTh}Summary of the values of $\mathrm{HDet}_{4}$, $S$ and $T$ invariants for the 15 transverse Ising model eigenstates $|\Psi_{k}\rangle$ with $0\leq k \leq 15$ as a function of $\lambda$ for $0\leq\lambda\leq 2/\sqrt{3}$. Functions $H(\alpha_\pm,\beta_\pm,\gamma_\pm)$, $S(\alpha_\pm,\beta_\pm,\gamma_\pm)$ and $T(\alpha_\pm,\beta_\pm,\gamma_\pm)$ are written in \eqref{eq:STHDet_Ising} and \eqref{eq:abc+-_Ising}.}
\begin{tabular}{@{}llll}
\br
\textbf{State} & $\mathbf{HDet_{4}}$ & $\mathbf{S}$ & $\mathbf{T}$ \\
\mr
$|\Psi_{0}\rangle,|\Psi_{15}\rangle$ & $H(\alpha_+,\beta_+,\gamma_+)$ & $S(\alpha_+,\beta_+,\gamma_+)$ & $T(\alpha_+,\beta_+,\gamma_+)$ \\
$|\Psi_{1}\rangle,|\Psi_{5}\rangle,|\Psi_{10}\rangle,|\Psi_{14}\rangle$, & 0 & $(2^6 3)^{-1}(1+\lambda^2)^{-2}$ & $(2^9 3^3)^{-1}(1+\lambda^2)^{-3}$ \\
$|\Psi_{2}\rangle,|\Psi_{13}\rangle$ & $H(\alpha_-,\beta_-,\gamma_-)$ & $S(\alpha_-,\beta_-,\gamma_-)$ & $T(\alpha_-,\beta_-,\gamma_-)$ \\
$|\Psi_{3}\rangle,|\Psi_{4}\rangle,|\Psi_{3}\rangle,|\Psi_{7}\rangle,|\Psi_{8}\rangle,|\Psi_{11}\rangle,|\Psi_{12}\rangle$ & 0 & 0 & 0\\
$|\Psi_{6}\rangle,|\Psi_{9}\rangle$ & 0 & $(2^6 3)^{-1}$ & $-(2^9 3^3)^{-1}$\\
\br
\end{tabular}
\end{footnotesize}
\end{table}

To illustrate this result, let us write explicitly an  eigenstate for each type of  behavior.
Let's start with eigenstates with zero $\mathrm{HDet}_{4}$. As the neighboring interaction is ruled by $\sigma^{x}$, the states are written in terms of the eigenvalues of $\sigma^{x}$, i.e. $|\rightarrow\rangle=(|\uparrow\rangle+|\downarrow\rangle)/\sqrt{2}$ and $|\leftarrow\rangle=(|\uparrow\rangle-|\downarrow\rangle)/\sqrt{2}$, where $|\uparrow\rangle$ and $|\downarrow\rangle$ are the eigenvectors of $\sigma^{z}$. For simplicity, we use the computational basis, i.e. $|0\rangle\equiv|\rightarrow\rangle$ and $|1\rangle\equiv|\leftarrow\rangle$.

An  example is given by one of the degenerated third excited states
\begin{eqnarray}
|\Psi_{3}\rangle= \frac{1}{\sqrt{2}}\left(-|0010\rangle+|1000\rangle\right)=-|\Psi^-\rangle_{13}|00\rangle_{24},
\end{eqnarray}
where $|\Psi^-\rangle=(|01\rangle-|10\rangle)/\sqrt{2}$, the rest of the degenerated states of this level behave analogously (see Appendix \ref{app:Ising}).\\

For the first excited state, $S$ and $T$ are non zero but $\mathrm{HDet}_{4}=0$:
\begin{eqnarray}
\fl |\Psi_{1}\rangle &=&\frac{1}{2\sqrt{(\lambda+\sqrt{\lambda'})^2+1}}\left(  \left(\lambda+\sqrt{\lambda'}\right)  \left\{ |0001\rangle+|0010\rangle+|0100\rangle+|1000\rangle\right\} \nonumber \right.\\ \fl &+&\left.|0111\rangle+|1011\rangle+|1101\rangle+|1110\rangle\right).
\end{eqnarray}
Observe that  this state is a combination of two $|W\rangle$-type states \eqref{eq:W}.
%
There are other states where $S \neq 0$ and $T  \neq 0$, 
 but $\mathrm{HDet}_{4}=0$, namely 
\begin{eqnarray}
|\Psi_{6}\rangle=\frac{1}{\sqrt{2}}\left(-|0011\rangle+|1100\rangle\right),\nonumber\\
|\Psi_{9}\rangle=\frac{1}{\sqrt{2}}\left(-|0101\rangle+|1010\rangle\right).
\label{eq:MaxEnt2}
\end{eqnarray}
These states have the same values of $S$ and $T$ as the GHZ states and are  not separable in any bipartition  but they entangle half of the system with the other half. In fact, they represent the two ways of  maximally entangle two spins in one direction with the other two in the opposite direction. If we define the states $|\rightrightarrows\rangle\equiv|00\rangle$ and $|\leftleftarrows\rangle\equiv|11\rangle$, then $|\Psi_{6}\rangle=\frac{1}{\sqrt{2}}\left(-|\rightrightarrows\rangle_{12}|\leftleftarrows\rangle_{34}+|\leftleftarrows\rangle_{12}|\rightrightarrows\rangle_{34}\right)$ and $|\Psi_{9}\rangle=\frac{1}{\sqrt{2}}\left(-|\rightrightarrows\rangle_{13}|\leftleftarrows\rangle_{24}+|\leftleftarrows\rangle_{13}|\rightrightarrows\rangle_{24}\right)$, which are $|\Psi^-\rangle$ states.

There are four states with non-zero $\mathrm{HDet}_{4}$: ground state and second, thirteenth and fifteenth excited states. The corresponding functions of $S$, $T$ and $H$ shown in table \ref{Tab:IsingTh} are
\begin{eqnarray}
S(\alpha,\beta,\gamma)&=&\frac{\Gamma(\alpha,\beta,\gamma)}{12\mathcal{N}(\alpha,\beta,\gamma)^2},\nonumber\\
T(\alpha,\beta,\gamma)&=&\frac{\left(4 \beta^2 (\alpha + \gamma^2)-(\alpha- \gamma^2)^2\right)\left(\Gamma (\alpha,\beta,\gamma)-768 \alpha \beta^4\gamma^2\right)}{216\mathcal{N}(\alpha,\beta,\gamma)^3},\nonumber\\
H(\alpha,\beta,\gamma)&=& S(\alpha,\beta,\gamma)^3-27T(\alpha,\beta,\gamma)^2,
\label{eq:STHDet_Ising}
\end{eqnarray}
where $\Gamma(\alpha,\beta,\gamma)= \alpha^2 (\alpha-4 \beta^2)^2 -4 \alpha(\alpha^2 -2\alpha \beta^2 -56 \beta^4) \gamma^2 +2 (3 \alpha^2 + 4 \alpha \beta^2 + 8\beta^4)\gamma^4 -4 (\alpha + 2 \beta^2) \gamma^6 + \gamma^8$ and $
\mathcal{N}(\alpha,\beta,\gamma)=(1 + \alpha^2 + 4\beta^2 + 2\gamma^2)^2$, which is the fourth power of the norm of these states as a function of $\alpha$, $\beta$ and $\gamma$ parameters. 
These parameters are functions of $\lambda$ and for the ground state and second excited state are
\begin{align}
\alpha_{\pm}&= \frac{1}{\lambda}\left(2 \lambda^3 +\sqrt{2}\lambda^2\sqrt{\lambda' \pm\sqrt{\lambda''}}-\sqrt{2}\sqrt{\lambda' \pm\sqrt{\lambda''}} \left(1 \mp \sqrt{\lambda''}\right) - \lambda\left(1 \mp2\sqrt{\lambda''}\right)\right),\nonumber\\
\beta_{\pm}&= \lambda+\frac{1}{\sqrt{2}}\sqrt{\lambda' \pm\sqrt{\lambda''}}, \nonumber\\
\gamma_{\pm}&= 1 + \frac{\sqrt{2} \lambda}{\sqrt{\lambda' \pm\sqrt{\lambda''}}}.
\label{eq:abc+-_Ising}
\end{align}
The ground state and the second excited state in terms of these parameters become
\begin{equation} 
|\Psi_{\pm}\rangle \propto \alpha_{\pm}|0000\rangle+ \beta_{\pm}\left(|0011\rangle+|0110\rangle+|1001\rangle+ |1100\rangle\right) +\gamma_{\pm}\left(|0101\rangle+|1010\rangle\right)+|1111\rangle,
\label{eq:36} 
\end{equation}
where $|\Psi_{+}\rangle\equiv|\Psi_{0}\rangle$ and $|\Psi_{-}\rangle\equiv|\Psi_{2}\rangle$. Eq.~\eqref{eq:36} shows  how rich is the quadripartite entanglement in these states. They contain all entanglement forms seen previously: part of the state is separable into two subsystems, other part of the state entangles maximally two spins in $|0\rangle$ state with two spins in $|1\rangle$ state and also contain the states with all spins aligned.

Figure \ref{Fig:IsThHdet} shows  $\mathrm{HDet}_{4}$ for  the ground state and the second excited state. 
Both curves  have peaks at  different values of $\lambda$: the ground state $\mathrm{HDet}_{4}$ 
peaks at $\lambda\sim 0.8$, close to the critical point, which for chain of $n=4$ sites is $\lambda\simeq 0.7$, while the $\mathrm{HDet}_{4}$ 
of the second excited state peaks at $\lambda\sim 1.2$, where it is not the second excited state anymore, 
as $|\Psi_{2}\rangle$ intersects with $|\Psi_{3}\rangle$ at $\lambda=2/\sqrt{3}\sim 1.15$. 
The order of magnitude of the peaks  are  also different: when the ground state has  $\mathrm{HDet}_{4} \propto 10^{-16}$,
the second excited state has  $\mathrm{HDet}_{4} \propto 10^{-9}$, 
the mean value of $\mathrm{HDet}_{4}$ for a random state. Moreover, the excited state's  peak  is broader than the ground state's  peak. 
Then, even both states have the same analytic  structure, the differences in the coefficients of the wave function lead to 
a difference of seven orders between the two $\mathrm{HDet}_{4}$.


\begin{figure}[t!]
\centering
\includegraphics[width=0.6\textwidth]{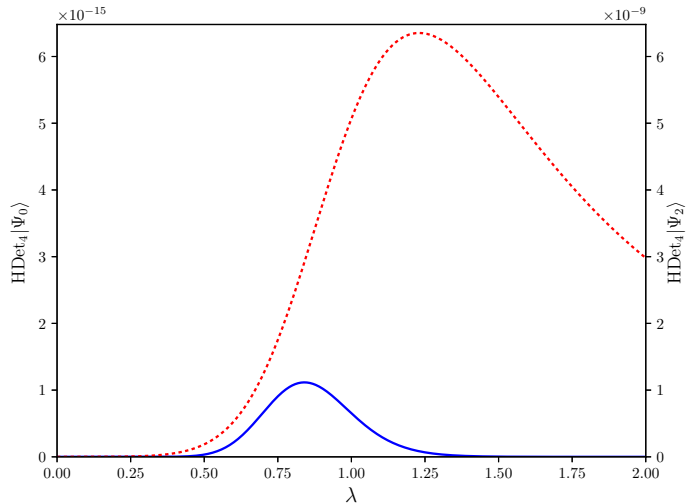}
\caption{$\mathrm{HDet}_{4}$ for the ground state $|\Psi_{0}\rangle$ (left axis, blue solid curve) 
and second excited state $|\Psi_{2}\rangle$ (right axis, dotted red curve) of the Ising model of $n=4$ spins as a function of the $\lambda$ field parameter. The $\mathrm{HDet}_{4}$ of the second excited state is seven orders of magnitude greater than the ground state's.}
\label{Fig:IsThHdet}
\end{figure}




\section{The Heisenberg XXZ Model \label{sec:XXZ}}

The XXZ model is a generalization of Heisenberg model
\begin{eqnarray}
H_{XXZ}=\sum_{i=1}^n \left( \sigma_{i}^{x}\sigma_{i+1}^{x}+\sigma_{i}^{y}\sigma_{i+1}^{y}+ \Delta \sigma_{i}^{z}\sigma_{i+1}^{z} \right) 
\label{eq:XXZ}
\end{eqnarray}
with the anisotropy parameter $\Delta$.

This model is critical in the region $\Delta\in(-1,1]$, known as the $XY$ phase \cite{Heisenberg}. Its entropy scales following a conformal scaling law with a central charge $c=1$, so it belongs to a different universality class than the Ising model. For $\Delta>1$ the system is in the N\'eel phase and for $\Delta<-1$ in the ferromagnetic phase. Then, this model present two quantum phase transitions, at $\Delta=1$ and at $\Delta=-1$. The first one is a Kosterlitz-Thouless where the gap scales as $\rme^{-\pi^2/2\sqrt{2(\Delta-1)}}$ for $\Delta$ slightly larger than one \cite{CP04}. The second transition at $\Delta = -1$ belongs to the Dzhaparidze-Nersesyan-Pokrovsky-Talapov 
universality class \cite{DN78,PT79}, where the entropy scales as $S \simeq \frac{1}{2} \log L$ at exactly $\Delta = -1$ \cite{EntXXZ}.  

\subsection{Eigenstates}

We diagonalize the XXZ Hamiltonian with $n=4$ spins and periodic boundary conditions. The energies as a function of $\Delta$ are
\begin{equation}
\left\lbrace -4,-4,4, 4, 0, 0, 0, 0, 0, 0, 0,  -4\Delta , 4\Delta,    4\Delta,-2\left(\Delta -\sqrt{8+\Delta^2}\right), -2\left(\Delta +\sqrt{8+\Delta^2}\right) \right\rbrace ,
\end{equation}
The order of the levels will depend on the value of $\Delta$. For $\Delta<-1$, the ground state is degenerate and correspond to the states with all spins aligned (ferromagnetic phase). 
For $\Delta>-1$ the ground state is unique and  has energy $-2(\Delta+\sqrt{8+\Delta^2})$. At the isotropic
point $\Delta =1$, it describes a 
resonating valence bound state (see below). 

The expressions of  $S$,  $T$ and  $\mathrm{HDet}_{4}$ for the states obtained after the diagonalization are summarized in table \ref{Tab:XXZ} 
(see \ref{app:XXZ} for details and the effects on degeneracy). In all states $\mathrm{HDet}_{4}$ is zero either because $S$ and $T$
vanish, or because they cancel each other in $\mathrm{HDet}_{4} = S^3 - 27 T^2$. We are using here the computational basis to describe the spin states written in the $\sigma_z$ basis, i.e. $|0\rangle\equiv|\uparrow\rangle$ and $|1\rangle\equiv|\downarrow\rangle$.

There are three types of states that lead to null $S$ and $T$ invariants. As in the  Ising model, there are states separable into two subsystems. For example, one of the states with zero energy  can be written as
\begin{eqnarray}
\frac{1}{\sqrt{2}}\left(|0111\rangle-|1101\rangle\right)=|\Psi^-\rangle_{13}|11\rangle_{24}.
\end{eqnarray}

Other type are the product states, of course. 
There are two of them in the XXZ spectrum:  $|0000\rangle$ and  $|1111\rangle$, where all spins are aligned, with an energy $4\Delta$. Both correspond to the ground states for $\Delta<-1$ and the most  excited states for $\Delta> 1$.

Finally, the third type of states are $W$-like. For example, one of the states with energy 4:
\begin{eqnarray}
\frac{1}{2}\left(|0111\rangle +|1011\rangle +|1101\rangle +|1110\rangle\right)
\end{eqnarray}

\begin{table}[t!]
\caption{\footnotesize\label{Tab:XXZ}$S$, $T$ and $\mathrm{HDet}_{4}$  of XXZ model for states obtained after the Hamiltonian diagonalization. All states lead to zero $\mathrm{HDet}_{4}$ and only four states have $S$ and $T$ invariants different from zero. The values can change in case of degeneracy, as it is explained in Appendix \ref{app:XXZ}. 
}
\begin{indented}
\begin{footnotesize}
\item[]\begin{tabular}{@{}llll}
\br
\textbf{Energy} & $\mathbf{S}$ & $\mathbf{T}$ & $\mathbf{HDet_{4}}$ \\
\mr
$-4(2)$, 4(2), 0(6), 4$\Delta$(2) & 0 & 0 & 0 \\
0, $-4\Delta$ &  $1/(2^6\cdot 3)$ & $-1/(2^9\cdot 3^3)$ & 0 \\
$-2\left(\Delta-\sqrt{8+\Delta^2}\right)$ & $S_{+}$ & $T_{+}$ & 0\\
$-2\left(\Delta+\sqrt{8+\Delta^2}\right)$ & $S_{-}$ & $T_{-}$ & 0\\
\br
\end{tabular}
\end{footnotesize}
\end{indented}
\end{table}

Only four energies have $S$ and $T$ different from zero. Two of them, with energies 0 and $-4\Delta$, correspond with the two ways of maximally entangle two sets of spins in opposite directions. These are the same states of the Ising model but in $\sigma_z$ basis, i.e. $|\upuparrows\rangle\equiv|00\rangle$ and $|\downdownarrows\rangle\equiv|11\rangle$. Then, these states become $\frac{1}{\sqrt{2}}\left(-|\upuparrows\rangle_{12}|\downdownarrows\rangle_{34}+|\downdownarrows\rangle_{12}|\upuparrows\rangle_{34}\right)$ and $\frac{1}{\sqrt{2}}\left(-|\upuparrows\rangle_{13}|\downdownarrows\rangle_{24}+|\downdownarrows\rangle_{13}|\upuparrows\rangle_{24}\right)$. Both states have $S$ and $T$ constant and with the same value as in the Ising model case, i.e. $S=1/(2^6\cdot 3)$ and $T=-1/(2^9\cdot 3^3)$.

On the other hand, there are two states with $S$ and $T$ that depend on $\Delta$. 
One with energy $-2\left(\Delta+\sqrt{8+\Delta^2}\right)$ corresponds to the ground state for $\Delta> -1$:
\begin{eqnarray}
\fl \frac{1}{\mathcal{N}}\left(|0011\rangle +|0110\rangle +|1100\rangle +|1001\rangle -\frac{1}{2}\left(\Delta+\sqrt{8+\Delta^2}\right)\left(|0101\rangle+|1010\rangle\right)\right),
\label{eq:XXZ_gs}
\end{eqnarray}
where $\mathcal{N}=8 + \Delta(\Delta +\sqrt{8 + \Delta^2})$.
$S$ and $T$ are non zero as long as $\Delta\neq 1$. When $\Delta=1$ it becomes a resonating valence bound state, as it is shown in the next subsection. The other state has energy $-2\left(\Delta-\sqrt{8+\Delta^2}\right)$ and corresponds to the state with higher energy for $\Delta<1$:
\begin{eqnarray}
\fl \frac{1}{\mathcal{N'}}\left(|0011\rangle +|0110\rangle +|1100\rangle +|1001\rangle -\frac{1}{2}\left(\Delta-\sqrt{8+\Delta^2}\right)\left(|0101\rangle+|1010\rangle\right)\right).
\label{eq:XXZ_e15}
\end{eqnarray}
where $\mathcal{N'}=8 + \Delta(\Delta -\sqrt{8 + \Delta^2})$.
This state has $S$ and $T$ different from zero as long as $\Delta\neq -1$. The expressions for the invariants of these two states are
\begin{eqnarray}
S_{\pm}&=&\frac{1}{2^8\cdot 3}\frac{\left(\Delta\pm\sqrt{8+\Delta^2}\right)^4\left(4-\Delta\left(\Delta\mp\sqrt{8+\Delta^2}\right)\right)^2}{\left(8+\Delta\left(\Delta\pm\sqrt{8+\Delta^2}\right)\right)^4}, \\
T_{\pm}&=&\frac{1}{2^{12}\cdot 3^3}\frac{\left(\Delta\pm\sqrt{8+\Delta^2}\right)^6\left(4-\Delta\left(\Delta\mp\sqrt{8+\Delta^2}\right)\right)^3}{\left(8+\Delta\left(\Delta\pm\sqrt{8+\Delta^2}\right)\right)^6},
\label{eq:ST_XXZ}
\end{eqnarray}
and are shown in figure \ref{Fig:XXZ}. Invariants for these two states seem to be sensible to the transition points $\Delta=1$ and $\Delta=-1$, as each one become zero at one of these points.

\begin{figure}
\centering
\includegraphics[width=0.6\textwidth]{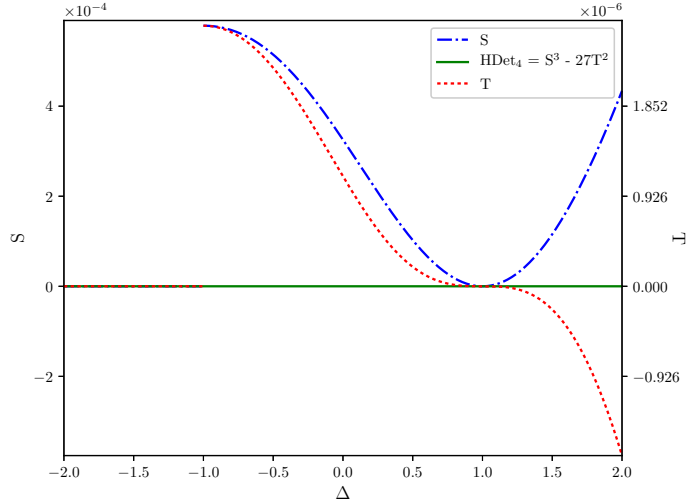}
\caption{$S$ and $T$ invariants of the ground state of $n=4$ XXZ spin chain. $\mathrm{HDet}_{4}$ is always zero but the $S$ and $T$ invariants are able to detect the transition points at $\Delta=-1$ and $\Delta=1$.}
\label{Fig:XXZ}
\end{figure}

The XXZ model for $\Delta=1$ is known as the XXX or isotropic Heisenberg model.
This  Hamiltonian is invariant under the rotation group, which allows for an easy derivation
of the spectrum and eigenstates. For $n=4$ spins, the Hamiltonian can be written as 
\begin{eqnarray}
\mathrm{H}_{XXX}&=&
 2\left[ s(s+1) - s_{13}(s_{13}+1)-s_{24}(s_{24}+1)\right],
\end{eqnarray}
where $s$ is the total spin and $s_{13}$ and $s_{24}$ are the total spins for particles 1 and 3,  and 2 and 4 respectively. 

Table \ref{Tab:XXZ_Spin} shows the different values of $s_{13}$, $s_{24}$ and $s$ and the corresponding energy. When the total spin is zero, the state is called a Resonating Valence Bound \cite{RVB}. There are two of them in Heisenberg spin chain:
\begin{eqnarray}
|\phi_{1}\rangle &=&
\frac{1}{2\sqrt{2}}\left(|0011\rangle +|0110\rangle +|1100\rangle +|1001\rangle -2\left(|0101\rangle+|1010\rangle\right)\right), \label{eq:phi1}\\
|\phi_{2}\rangle &=&\frac{1}{2}\left(|0011\rangle -|0110\rangle -|1001\rangle +|1100\rangle\right). \label{eq:phi2}
\end{eqnarray}
The first one corresponds to the ground state whereas the second is a lineal combination of the states with zero energy. Both have the property $S=T=0$. To check if this is a general property of the resonating valence bound states, we have checked that the state
\begin{eqnarray}
|\phi\rangle=\cos\theta|\phi_{1}\rangle + \rme^{\rmi\varphi}\sin\theta|\phi_{2}\rangle
\end{eqnarray}
also have $S$ and $T$ zero $\forall$ $\theta,\varphi$.

We can also check what is the effect of degeneracy on $\mathrm{HDet}_{4}$. Although all states of XXZ Hamiltonian have $\mathrm{HDet}_{4}=0$, linear combinations of states with the same energy could have $\mathrm{HDet}_{4}\neq 0$. A detailed analysis of degeneracy taking as example the Heisenberg model is explained in \ref{app:XXZ}.

\begin{table}[t!]
\begin{footnotesize}
\caption{{\footnotesize Energies for the $n=4$ Heisenberg model (XXZ model with $\Delta=1$) according to the total spin of their particles. When the total spin is zero, it is called a Resonating Valence Bound state.}}
\label{Tab:XXZ_Spin}
\begin{indented}
\item[]\begin{tabular}{C{1.5cm}C{1.2cm}C{1.2cm}C{1.2cm}}
\hline
\textbf{Energy} & $\mathbf{s_{13}}$ & $\mathbf{s_{24}}$ & $\mathbf{s}$ \\
\hline
$-8$ & 1 & 1 & 0 \\
$-4$ & 1 & 1 & 1 \\
$0$ & 0 & 1 & 1 \\
$0$ & 1 & 0 & 1 \\
$0$ & 0 & 0 & 0 \\
4 & 1 & 1 & 2 \\
\hline
\end{tabular}
\end{indented}
\end{footnotesize}
\end{table}

\subsection{Thermal state}

The $S$ and $T$ invariants for a thermal states of the XXZ spin chain with $n=4$ sites is computed using defintion of Eq.\eqref{eq:Th2}. 


Figure \ref{Fig:XXZth} shows $S$ invariant for a thermal state. As $\beta$ decreases, the amount of entanglement quantified by this invariant decreases until zero. As expected, the multipartite entanglement is lost at high temperatures.

\begin{figure}[t!]
\centering
\includegraphics[width=0.6\textwidth]{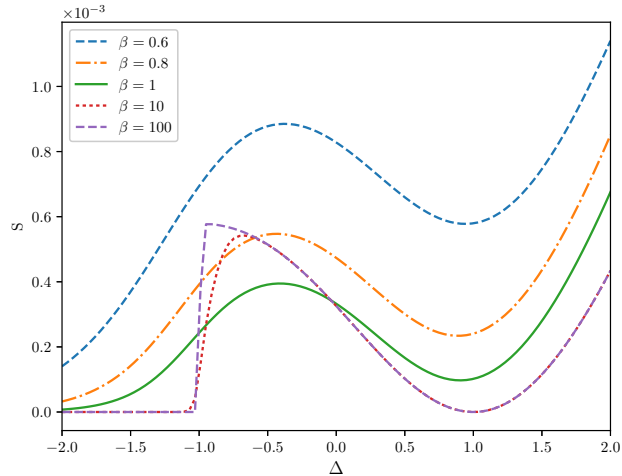}
\caption{$S$ invariant for the XXZ spin chain model as a function of $\Delta$ for different values of $\beta=1/(k_{B}T)$. The amount of entanglement quantified by the $S$ invariant tends to zero as the temperature increases, as expected.
}
\label{Fig:XXZth}
\end{figure}

\section{The generalized  Haldane-Shastry wave functions \label{sec:HS}}

The Haldane-Shastry (HS) model \cite{HS} describes a chain of equally spaced spin-$\frac{1}{2}$ particles in a circle with pairwise interactions inversely proportional to the square of the distance between the spins. The Hamiltonian of the HS model is
given by 
\begin{eqnarray}
H_{HS} =  \frac{  \pi^2}{n^2}  \sum_{i > j  }^n  \frac{ \mathbf{S}_i \cdot \mathbf{S}_j} { \sin^2   \frac{ \pi (i-j)}{n}} 
\label{hs1}
\end{eqnarray}
where $\mathbf{S}_i = \frac{1}{2} \mathbf{\bsigma}_i$ are spin $\frac{1}{2}$ matrices  acting at the site $i=1, \dots, n$. 
The ground state of this Hamiltonian can be written as \cite{CiracGerman}
\begin{eqnarray}
\psi(s_{1},\cdots ,s_{n})\propto \delta_{s}\rme^{\rmi  \frac{\pi}{2} \sum_{i:\mathrm{odd}}s_{i}}\prod_{i>j}^{n}\left\vert\sin\frac{\pi(i-j)}{n}\right\vert^{s_{i}s_{j}/2}.
\label{eq:HS_sin}
\end{eqnarray}
where the spin at the site $i=1, \dots, n$ is given by $s_i/2$ with 
$s_{i}=\pm 1$, and $\delta_{s}=1$ if $\sum_{i=1}^{n}s_{i}=0$ and zero otherwise. 
The latter condition implies  that the total third component of the spin vanishes, that is
$\langle \sum_i S^z_i \rangle =0$, but the HS state is also a singlet of the rotation group,
$\langle (\sum_i {\bf S}_i )^2 \rangle =0$.  The HS wave function has a huge overlap
with the ground state of the isotropic Heisenberg model. In fact,  for $n=4$ sites these two wave functions
are the same. The HS Hamiltonian belongs to the same
universality class as the isotropic Heisenberg model, which is described by
the Wess-Zumino-Witten model $SU(2)_1$  that has a central charge $c=1$.  

The wave function (\ref{eq:HS_sin})  was generalized in \cite{CiracGerman} to the following one 
\begin{eqnarray}
\psi(s_{1},\cdots ,s_{n})\propto \delta_{s}\rme^{\rmi  \frac{\pi}{2} \sum_{i:\mathrm{odd}}s_{i}}\prod_{i>j}^{n}\left\vert\sin\frac{\pi(i-j)}{n}\right\vert^{ \alpha s_{i}s_{j}}, 
\label{eq:HS_sin2}
\end{eqnarray}
and was used as a variational ansatz for the ground state of the XXZ model in the critical regime. 
The relation between the anisotropy parameter $\Delta$ and the parameter $\alpha$
was found to be $\Delta=-\cos(2\pi\alpha)$, with  $0 <  \alpha \leq \frac{1}{2}$,
corresponding to the critical region $-1 < \Delta \leq 1$ \cite{CiracGerman}. 
The cases $\alpha = 0, \frac{1}{4}$ provide the exact solution of the XXZ model for $\Delta =-1, 0$, 
while  $\alpha = \frac{1}{2}$, is the HS wave function (\ref{eq:HS_sin}). 

\begin{figure}[h!]
\centering
\includegraphics[width=0.55\textwidth]{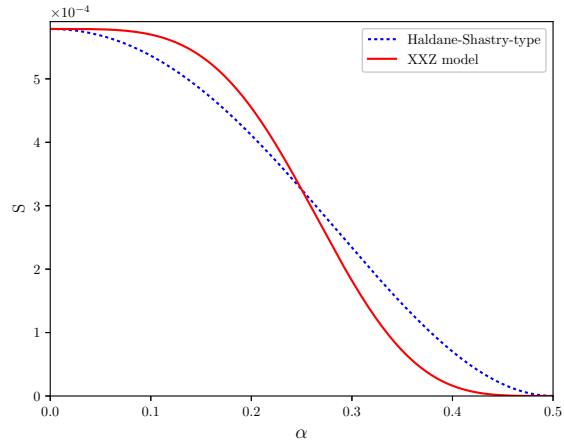}
\caption{Comparison of the $S$ invariant of the ground state of the XXZ model and the wave function (\ref{eq:HS_sin2}) 
for $n=4$ spins. Both wave functions coincide for $\Delta =-1, 0, 1$ that correspond to $\alpha =0, \frac{1}{4}, \frac{1}{2}$. } 
\label{Fig:S_alpha}
\end{figure}

\subsection{Ground state and $S$ and $T$ invariants}

In the ground state of the HS  model the total spin vanishes, that is  $\sum_{i=1}^{n}s_{i}=0$. For $n=4$ spins, the wave function becomes a superposition of the states
\begin{equation}
|\Psi\rangle=\frac{1}{\sqrt{1+3\cdot 4^{-2\alpha}+4^{-\alpha}}}\left(4^{-\alpha}(|0011\rangle+|0110\rangle+|1001\rangle+|1100\rangle)-(|0101\rangle+|1010\rangle)\right),
\label{eq:HScomp}
\end{equation}
where we have used the computational basis $|0\rangle(|1\rangle)$ to describe the spins $s_{i}=-1(+1)$. This type of wave function have $\mathrm{HDet}_{4}=0$ as a consequence of the cancellation of $S$ and $T$ invariants
\begin{eqnarray}
S_{HS}&=&\frac{4^{4\alpha-3}\left(16^\alpha-4\right)^2}{3(2+16^\alpha)^4},\nonumber\\
T_{HS}&=&-\frac{8^{4\alpha-3}\left(16^\alpha-4\right)^3}{27(2+16^\alpha)^6}.
\label{eq:ST_HS}
\end{eqnarray}
Thus, as in the  XXZ model, we shall study the $S$ and $T$ behaviors instead of $\mathrm{HDet}_{4}$ which vanishes identically.

Figure \ref{Fig:S_alpha} shows the $S$ invariant as a function of $\alpha$ parameter. As expected, it matches with the XXZ $S$ invariant at $\alpha=0,\frac{1}{4},\frac{1}{2}$. Also, $\alpha=\frac{1}{4}$ is the inflexion point: for $\alpha<\frac{1}{4}$, $S_{XXZ}>S_{HS}$ whereas for $\alpha>\frac{1}{4}$, $S_{XXZ}<S_{HS}$.

\subsection{Dimerized wave function}

We can modify the interaction strength between the spins introducing a new parameter $\delta$,
and the  wave function 
\begin{eqnarray}
\psi_{\delta}(s_{1},\cdots ,s_{n})&\propto& \delta_{s}\rme^{\rmi \frac{\pi}{2} \sum_{i:\mathrm{odd}}s_{i}}\prod_{i>j}^{n}\left\vert 2\sin\left(\theta_{i}-\theta_{j}\right)\right\vert^{\alpha s_{i}s_{j}},
\label{eq:HSdelta}
\end{eqnarray}
where $\theta_{j}=\pi/n\left(j+\delta(-1)^j\right)$ for $j=1,\cdots ,n$. If  $\delta=0$ 
this wave function becomes (\ref{eq:HS_sin}). 

The wave function and $S$ and $T$ invariants become
\begin{eqnarray}
|\Psi_{\delta}\rangle &\propto & a_{1}\left(|0011\rangle+|1100\rangle\right) +a_{2}\left(|0101\rangle+|1010\rangle\right) +a_{3}\left(|0110\rangle+|1001\rangle\right),\nonumber\\
S&=&\frac{\left(a_{1}^4+\left(a_{2}^2-a_{3}^2\right)^2-2a_{1}^2\left(a_{2}^2+a_{3}^2\right)\right)^2}{192\left(|a_{1}|^2+|a_{2}|^2+|a_{3}|^2\right)^4},\nonumber\\
T&=&-\frac{\left(a_{1}^4+\left(a_{2}^2-a_{3}^2\right)^2-2a_{1}^2\left(a_{2}^2+a_{3}^2\right)\right)^3}{13824\left(|a_{1}|^2+|a_{2}|^2+|a_{3}|^2\right)^6},
\end{eqnarray}
where
\begin{eqnarray}
a_{1}&=&-2^{-\alpha}\left|\frac{\cos\left(\pi(3+2\delta)/4\right)}{\cos(\pi\delta/2)-\sin(\pi\delta/2)}\right|^{2\alpha},\nonumber\\
a_{2}&=&|\cos(\pi\delta)|^{-2\alpha},\nonumber\\
a_{3}&=&-4^{-\alpha}\left|1-\frac{2}{1+\tan(\pi\delta/2)}\right|^{2\alpha}.
\end{eqnarray}

\begin{figure}[t!]
\centering
\begin{subfigure}[b]{0.45\textwidth}
\centering
\includegraphics[width=\textwidth]{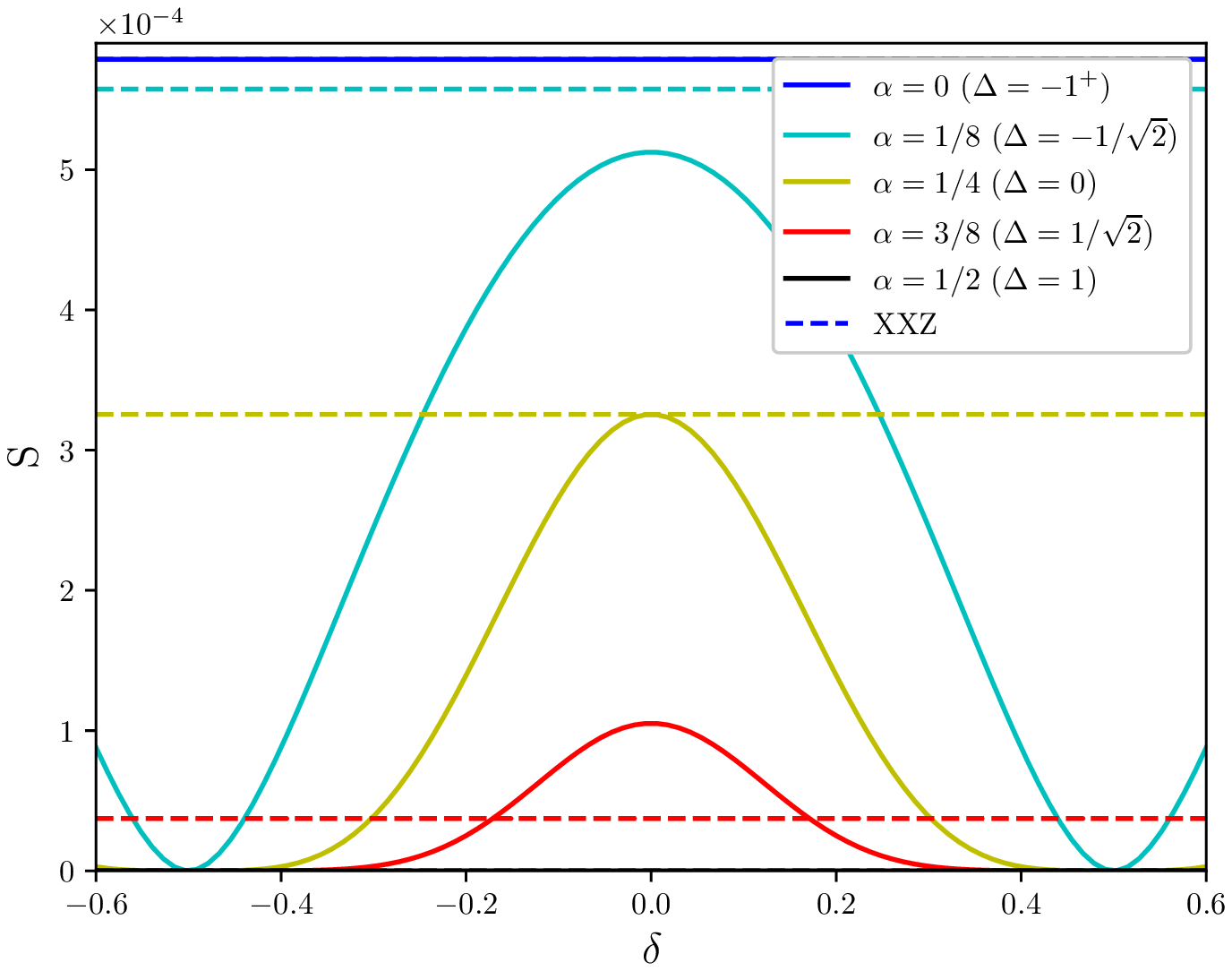}
\end{subfigure}%
\begin{subfigure}[b]{0.45\textwidth}
\centering
\includegraphics[width=1.3\textwidth]{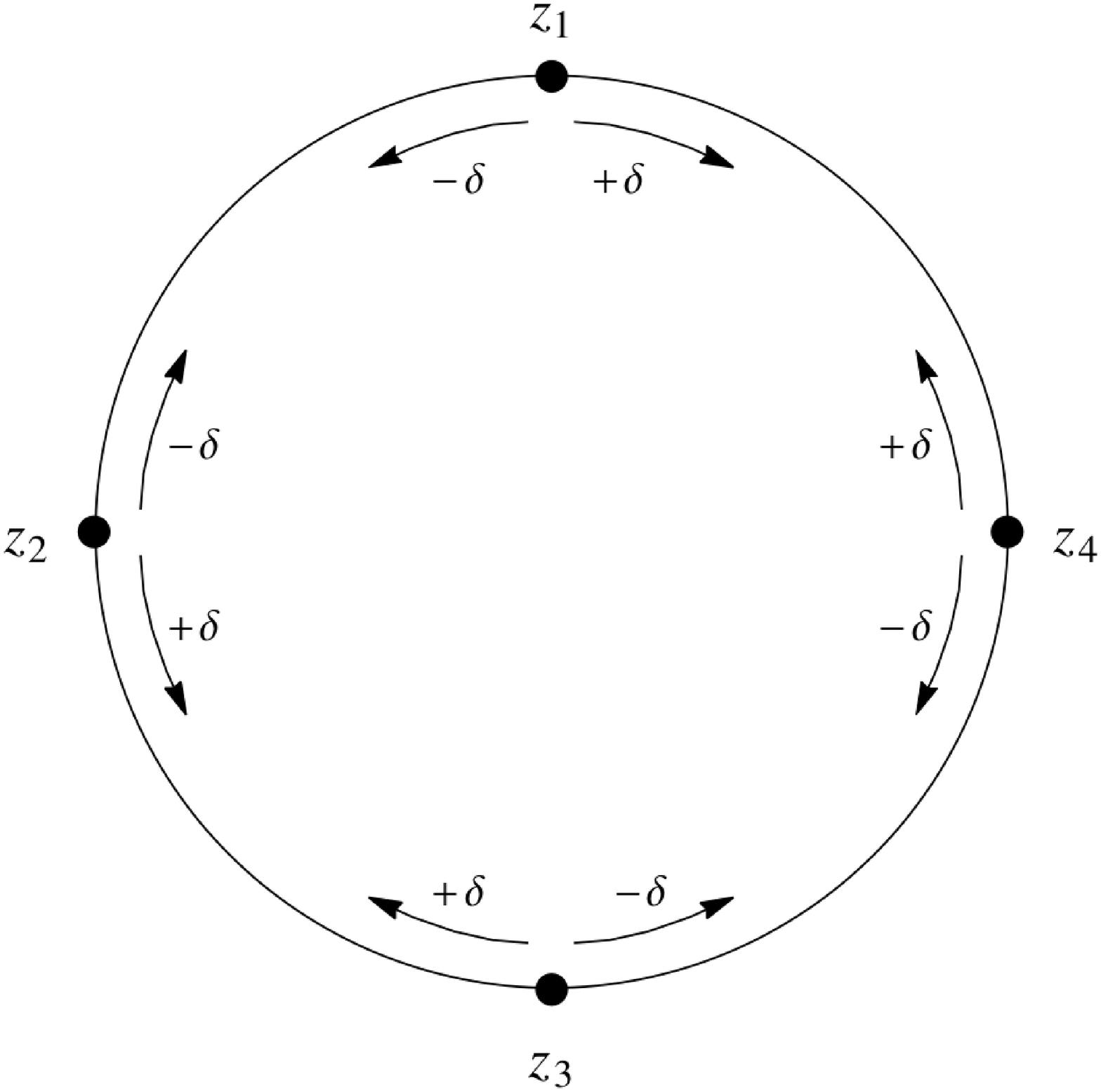}
\end{subfigure}
\caption{Left: $S$ invariant as a function of $\delta$ parameter for different values of $\alpha$.
Right: Diagrammatic representation of the $n=4$ Haldane-Shastry spin chain with the dimerization parameter $\delta$. For $\delta>0$ spins 1 and 4 and 2 and 3 are attracted each other, while for $\delta<0$ the attraction is between spins 1 and 2 and 3 and 4. For $|\delta|=\frac{1}{2}$ two consecutive spins are at the same position and the ground state is divided into two singlet states (dimer) and, as a consequence, $S$ and $T$ invariants become zero}
\label{Fig:S_delta}
\end{figure}

Figure \ref{Fig:S_delta} left shows the $S$ invariant as a function of $\delta$ parameter for different $\alpha$ values. It matches with XXZ model at $\alpha=0,\frac{1}{2}$ and shows a periodicity $S(\alpha,\delta)=S(\alpha,\delta\pm 1)$. Its maximum are located at $\delta=\pm m$ and its minimum at $\delta=\pm \frac{m}{2}$ for integer $m$. Moreover, maximum for $\alpha=\frac{1}{4}$ matches with $S$ invariant for the XXZ model at $\Delta=0$, as expected. In fact, it is enough to consider $\delta\in[-\frac{1}{2},\frac{1}{2}]$. We can write the wave function \eqref{eq:HSdelta} using the complex numbers $z_{j}=e^{2 i \theta_{j}}$. Then, $z_{j}$ correspond with the position of local spins, so at $\delta=\frac{1}{2}(-\frac{1}{2})$, spins 1 and 4 (1 and 2) and 2 and 3 (3 and 4) are at the same position and the state is a product of two singlets, i.e. dimer, as it is shown diagrammatically in right of figure \ref{Fig:S_delta}. Then, the state of four spins is separable into two subsystems and $S$ and $T$ become zero. A diagrammatic representation of the effect of $\delta$ is shown in figure \ref{Fig:S_delta} right.


\section{Conclusions \label{sec:conclusions}}

In  this work we have studied the quadripartite entanglement of several quantum states 
of four spins $\frac{1}{2}$, in particular in the following models:  Ising with a transverse field, XXZ  and Haldane-Shastry type model.  
We have used as a figure of merit the Schl\"afli hyperdeterminant  $\mathrm{HDet}_{4}$ \cite{Schlafli}, which  is an extension of the  $2\times 2\times 2\times 2$ dimensional 
Cayley's hyperdeterminant \cite{Cayley},
constructed from two polynomial  invariants $S$ and $T$,
as $\mathrm{HDet}_{4} = S^3 - 27 T^2$.
The latter quantities provide a more refined  characterization of the quadripartite entanglement, particularly in those
cases where $\mathrm{HDet}_{4}$ vanishes. We have also studied the $\mathrm{HDet}_{4}$ values of randomly distributed pure states.
An overview of the results is shown in the $S-T$ diagram plotted  in  figure \ref{Fig:landscape}. 

 \begin{figure}[h!]
\centering
\includegraphics[width=0.8\textwidth]{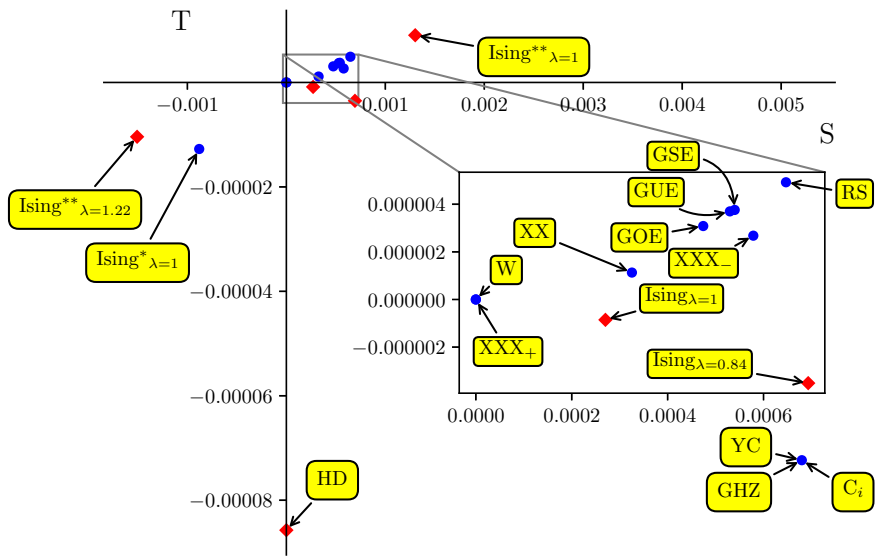}
\caption{\emph{Entanglement landscape.} $S-T$ plot for several wave functions analyzed in this work. For the  Ising model we plot ground state, $1^{st}$ and $2^{nd}$ excited states, denoted respectively with $*$ and $**$. For the  XXZ model we plot $\Delta=0$, that is the XX model, and $\Delta=\pm 1$ ($XXX_\pm$). RS stands for a typical random state and GOE, GUE and GSE for typical ground states of  random matrix Hamiltonians. Due to relation \eqref{eq:ST}, some states have zero $\mathrm{HDet}_{4}$, then we indicate with red diamond points the states with $\mathrm{HDet}_{4}\neq 0$.}
\label{Fig:landscape}
\end{figure}

We found that $\mathrm{HDet}_{4}$ is sensible to different priors on such random states. In particular, we analyzed flat and Haar random distributed state coefficients and the ground states of random matrices: GUE, GSE and GOE. The mean value of $\mathrm{HDet}_{4}$ is different between flat and Haar distribution, and the last has similar value as GUE and GSE random matrices.


 
For the Ising model, we found that ground state $\mathrm{HDet}_{4}$ shows a pronounced peak at $\lambda=0.84$, that lies near the critical point of the model for $n=4$ spins, located at $\lambda\simeq 0.7$. This small deviation of the peak from the critical value for infinite chains $\lambda=1$ can be attributed  to finite size effects.

The XXZ model exhibit vanishing values of  $\mathrm{HDet}_{4}$ for all non-degenerate states.
This fact  is due to an exact cancellation between  the $S$ and $T$ terms in the equation $\mathrm{HDet}_{4} = S^3 - 27 T^2$. 
In the whole critical regime $- 1 < \Delta \leq 1$, one has $S \geq 0$, and
there is a discontinuity at the point $\Delta =-1$. On the other hand, in the antiferromagnetic
regime $\Delta >1$, one has that $S < 0$. Hence, the invariant $S$ is able to distinguish
between the different  phases of this model. 

The results obtained for the  Haldane-Shastry type model are similar 
to  those of the  XXZ model
in the critical regime.
We also introduce a dimerization factor $\delta$ and study the multipartite entanglement as a function of this coefficient. The result shows that $S$ and $T$ invariants are maximum when $\delta=0$ and zero when $\delta=\frac{1}{2}$, which corresponds to two consecutive spins at the same physical position: the state becomes a product state of two singlets (dimer).


In summary, we have shown that  Cayley hyperdeterminant is a useful tool to characterize
the multipartite entanglement in several wave functions. In the case of random distributed states, it is sensible to the prior used. This analysis can be extended to other priors than the ones used in this work. In the analysis of states with 4 spins $\frac{1}{2}$, it is able to detect
phase transitions 
even for  such a small number of degrees of freedom. 
A direct extension to higher values of the spin or  more sites seems at the moment out of reach, but  
it suggests new tools to characterize multipartite entanglement along this direction.


\vspace{0.5 cm}

\ack{} 


AC  and JIL   acknowledge support from the grant FIS2015-69167-C2-2-P,  and
GS acknowledges the support from the grants 
FIS2015-69167-C2-1-P, QUITEMAD+ S2013/ICE-2801 and SEV-2016-0597 of the 
``Centro de Excelencia Severo Ochoa" Programme.

\begin{appendix}

\section{Hyperdeterminant for the nine classes of quadripartite states\label{app:HDet}}

In this appendix, we present the results for the computation of $\mathrm{HDet}_{4}, S$ and $T$ invariants for the classification of quadripartite entangled states defined by Verstraete {\sl et al.} in \cite{Verstraete}.

There is only one family of states with $\mathrm{HDet}_{4}$ different from zero:
\begin{eqnarray}
\fl G_{abcd}&=&\frac{a+d}{2}\left(|0000\rangle+|1111\rangle\right)+\frac{a-d}{2}\left(|0011\rangle+|1100\rangle\right) \nonumber \\  \fl &+& \frac{b+c}{2}\left(|0101\rangle+|1010\rangle\right)+\frac{b-c}{2}\left(|0110\rangle+|1001\rangle\right),
\end{eqnarray}
whose  values for $S$, $T$ and $\mathrm{HDet}_{4}$ are given by
\begin{align}
S=&\frac{1}{12}\left((b^2 - c^2)^2 (a^2 - d^2)^2 + (a^2 - b^2) (b^2 - c^2) (a^2 -d^2) (c^2 - d^2) + (a^2 - b^2)^2 (c^2 - d^2)^2\right),\nonumber\\
 T=&\frac{1}{1728}\left((a c + b d)^2 + (a b + c d)^2-2 (b c + a d)^2\right)  \nonumber \\ 
& \times  \Big( \left((a c + b d)^2 + (a b + c d)^2-2 (b c + a d)^2 \right)^2 - 9 (b - c)^2 (b + c)^2 (a - d)^2 (a + d)^2\Big), 
\nonumber\\
\mathrm{HDet}_{4}=&\frac{1}{256}(a^2 - b^2)^2 (a^2 - c^2)^2 (b^2 - c^2)^2 (a^2 - d^2)^2 (b^2 - d^2)^2 (c^2 - d^2)^2,
\end{align}

Notice that if two parameters are equal, $\mathrm{HDet}_{4}$ become zero. States \eqref{eq:XXZ_gs} and \eqref{eq:XXZ_e15} of XXZ model are of this type: correspond to the cases where  $a=-d$, which makes $S$ and $T$ proportional to $(a^2 - b^2)(a^2 - c^2)$. For $\Delta=1$, $a=-b$ in state \eqref{eq:XXZ_gs} and for $\Delta=-1$, $a=c$ in state of \eqref{eq:XXZ_e15}.

There are three families of states with $S$ and $T$ non zero in general. These are
$L_{abc_{2}}=\frac{a+b}{2}\left(|0000\rangle+|1111\rangle\right)+\frac{a-b}{2}\left(|0011\rangle+|1100\rangle\right)+c\left(|0101\rangle+|1010\rangle\right)+|0110\rangle$
with
\begin{eqnarray}
S&=&\frac{1}{12}(a^2 - c^2)^2 (c^2 - b^2)^2, \qquad T=\frac{1}{216}(a^2 - c^2)^3 (c^2 - b^2)^3, 
\end{eqnarray}
$
L_{a_{2}b_{2}}=a\left(|0000\rangle+|1111\rangle\right)+b\left(|0101\rangle+|1010\rangle\right)+|0110\rangle+|0011\rangle
$
with
\begin{eqnarray}
S&=&\frac{1}{12}(a - b)^4 b^4, \qquad   T = -\frac{1}{216}(a - b)^6 b^6 
\end{eqnarray}
and
$
L_{a_{2}0_{3\bigoplus\bar{1}}}=a\left(|0000\rangle+|1111\rangle\right)+|0011\rangle+|0101\rangle+|0110\rangle,
$
with
\begin{eqnarray}
S&=& \frac{1}{12}a^8,   \qquad T= -\frac{1}{216}a^{12}.  
\end{eqnarray}
$\mathrm{HDet}_{4}$ is zero for  these states. Finally, the  families
\begin{eqnarray}
\fl L_{ab_{3}} &=& a\left(|0000\rangle+|1111\rangle\right)+\frac{a+b}{2}\left(|0101\rangle+|1010\rangle\right)+ \\ \fl && \frac{a-b}{2}\left(|0110\rangle+|1001\rangle\right)
+\frac{\rmi}{\sqrt{2}}\left(|0001\rangle+|0010\rangle+|0111\rangle+|1011\rangle\right),  \nonumber  \\
\fl L_{a_{4}}&=&a\left(|0000\rangle+|0101\rangle+|1010\rangle+|1111\rangle\right)+\rmi|0001\rangle+|0110\rangle-\rmi|1011\rangle \nonumber \\
\fl L_{0_{5\bigoplus\bar{3}}}&=&|0000\rangle+|0101\rangle+|1000\rangle+|1110\rangle,  \nonumber  \\ 
\fl L_{0_{7\bigoplus\bar{1}}}&=&|0000\rangle+|1011\rangle+|1101\rangle+|1110\rangle, \nonumber \\
\fl L_{0_{3\bigoplus\bar{1}}0_{3\bigoplus\bar{1}}}&=&|0000\rangle+|0111\rangle  \nonumber
\end{eqnarray}
have $S$ and $T$ equal to zero.

\section{Ising model eigenstates\label{app:Ising}}

We collect below the eigenvalues and eigenstates of an Ising spin chain with $n=4$ sites and the formulas  of 
$\mathrm{HDet}_{4},  S$ and $T$ invariants; all of them are written in table \ref{Tab:Ising} and are labeled for $\lambda<2/\sqrt{3}$. 
We used the basis  $|0\rangle$ and $|1\rangle$ of eigenstates of $\sigma_{x}$. The coefficients $\alpha$, $\beta$ and $\gamma$ 
 appearing  in table \ref{Tab:Ising} are
\begin{eqnarray}
\fl \alpha_{0\pm} &=& \frac{1}{\lambda}\left(2 \lambda^3 \pm\sqrt{2}\lambda^2\sqrt{\lambda' +\sqrt{\lambda''}}\mp\sqrt{2}\sqrt{\lambda' +\sqrt{\lambda''}} \left(1 - \sqrt{\lambda''}\right) - \lambda\left(1 -2\sqrt{\lambda''}\right)\right), \label{B1}  \\ \fl
\alpha_{2\pm}&=& \frac{1}{\lambda}\left(2 \lambda^3 \pm\sqrt{2}\lambda^2\sqrt{\lambda' -\sqrt{\lambda''}}\mp\sqrt{2}\sqrt{\lambda' -\sqrt{\lambda''}}\left(1 + \sqrt{\lambda''}\right) - \lambda\left(1 +2\sqrt{\lambda''}\right)\right),\nonumber\\ \fl
\beta_{0\pm}&=& \lambda\pm\frac{1}{\sqrt{2}}\sqrt{\lambda' +\sqrt{\lambda''}},  \qquad 
\beta_{2\pm} =  \lambda\pm\frac{1}{\sqrt{2}}\sqrt{\lambda' -\sqrt{\lambda''}}
\nonumber \\ \fl
\gamma_{0\pm}&=& 1 \pm \frac{\sqrt{2} \lambda}{\sqrt{\lambda'} +\sqrt{\lambda''}},   \qquad 
\gamma_{2\pm} =  1 \pm \frac{\sqrt{2} \lambda}{\sqrt{\lambda'} -\sqrt{\lambda''}}
 \nonumber 
\label{eq:Isingvar}
\end{eqnarray}
where $\lambda'=1+\lambda^2$ and $\lambda''=1+\lambda^4$.

The states where  $S=T=0$ can be factorized into two subsystems:
\begin{eqnarray}
|\Psi_{3}\rangle &=& -|\psi\rangle_{13}|00\rangle_{24},  \quad  |\Psi_{4}\rangle = -|00\rangle_{13}|\psi\rangle_{24}, 
\quad |\Psi_{7}\rangle = -|01\rangle_{13}|\psi\rangle_{24} 
\\
|\Psi_{8}\rangle &=& -|\psi\rangle_{13}|01\rangle_{24},  \quad   |\Psi_{11}\rangle = -|11\rangle_{13}|\psi\rangle_{24},
\quad  |\Psi_{12}\rangle = -|\psi_{13}\rangle|11\rangle_{24}
 \nonumber 
\end{eqnarray}

\begin{table}[t!]
\begin{footnotesize}
\caption{\footnotesize
$S, T$  and $\mathrm{HDet}_{4}$ invariants  for the  Ising spin chain states with 4 sites, where $\lambda'=1+\lambda^2$. The states are ordered from the ground state to the $15^{th}$ excited state with the corresponding energies shown in \eqref{eq:Ising_energy}. The expressions for $\alpha$, $\beta$ and $\gamma$ are written in\eqref{B1}, and those of $S_{0,2}$,$T_{0,2}$ and $H_{0,2}$ in \eqref{eq:STHDet_Ising}.}
\label{Tab:Ising}
\begin{tabular}{C{9cm}ccc}
\hline
\textbf{State}  & $\mathbf{S}$ & $\mathbf{T}$  & $\mathbf{HDet}_{\mathbf{4}}$ \\
 \hline
$|\Psi_{0}\rangle\propto\alpha_{0+}|0000\rangle+ \beta_{0+}(|0011\rangle+|0110\rangle+|1001\rangle+|1100\rangle )+\gamma_{0+}(|0101\rangle+|1010\rangle)+|1111\rangle $ & $S_{0}$ &$T_{0}$ & $H_{0}$\\
$|\Psi_{1}\rangle\propto\left(\lambda+\sqrt{\lambda'}\right)(|0001\rangle+|0010\rangle+|0100\rangle+|1000\rangle)+|0111\rangle+|1011\rangle+|1101\rangle+|1110\rangle$ & $\left(2^6 3(\lambda')^{2}\right)^{-1}$ & $-\left(2^9 3^3(\lambda')^{3}\right)^{-1}$ & 0\\
$|\Psi_{2}\rangle\propto\alpha_{2+}|0000\rangle+ \beta_{2+}(|0011\rangle+|0110\rangle+|1001\rangle+|1100\rangle) +\gamma_{2+}(|0101\rangle+|1010\rangle)+|1111\rangle$ & $S_{2}$ &$T_{2}$ &$H_{2}$\\
 $|\Psi_{3}\rangle\propto-|0010\rangle + |1000\rangle$ & 0 & 0 & 0\\
 $|\Psi_{4}\rangle\propto -|0001\rangle+|0100\rangle$ & 0 & 0 & 0\\
$|\Psi_{5}\rangle\propto\left(\lambda+\sqrt{\lambda'}\right)(|0010\rangle-|0001\rangle-|0100\rangle+|1000\rangle)-|0111\rangle+|1011\rangle-|1101\rangle+|1110\rangle$ & $\left(2^6 3(\lambda')^{2}\right)^{-1}$ & $-\left(2^9 3^3(\lambda')^{3}\right)^{-1}$ & 0\\
 $|\Psi_{6}\rangle\propto-|0011\rangle+|1100\rangle$ & $\left(2^6 3\right)^{-1}$ & $-\left(2^9 3^3\right)^{-1}$ & 0\\
$|\Psi_{7}\rangle\propto-|0011\rangle+|0110\rangle$ & 0 & 0 & 0\\
 $|\Psi_{8}\rangle\propto-|0011\rangle+|1001\rangle$ & 0 & 0 & 0\\
 $|\Psi_{9}\rangle\propto -|0101\rangle+|1010\rangle$ & $\left(2^6 3\right)^{-1}$ & $-\left(2^9 3^3\right)^{-1}$ & 0\\
$|\Psi_{10}\rangle\propto\left(\lambda-\sqrt{\lambda'}\right)(|0001\rangle+|0010\rangle+|0100\rangle+|1000\rangle)+|0111\rangle+|1011\rangle+|1101\rangle+|1110\rangle$ & $\left(2^6 3(\lambda')^{2}\right)^{-1}$ & $-\left(2^9 3^3(\lambda')^{3}\right)^{-1}$ & 0\\

 $|\Psi_{11}\rangle\propto-|1011\rangle+|1110\rangle $ & 0 & 0 & 0\\
 $|\Psi_{12}\rangle\propto-|0111\rangle+|1101\rangle $ & 0 & 0 & 0\\ 
$|\Psi_{13}\rangle\propto\alpha_{2-}|0000\rangle+ \beta_{2-}(|0011\rangle+|0110\rangle+|1001\rangle+|1100\rangle)+\gamma_{2-}(|0101\rangle+|1010\rangle)+|1111\rangle$ & $S_{2}$ &$T_{2}$ & $H_{2}$\\ 

$|\Psi_{14}\rangle\propto\left(\lambda+\sqrt{\lambda'}\right)(|0010\rangle-|0001\rangle-|0100\rangle+|1000\rangle)-|0111\rangle+|1011\rangle-|1101\rangle+|1110\rangle$ & $\left(2^6 3(\lambda')^{2}\right)^{-1}$ & $-\left(2^9 3^3(\lambda')^{3}\right)^{-1}$ & 0\\
$|\Psi_{15}\rangle\propto\alpha_{0-}|0000\rangle+ \beta_{0-}(|0011\rangle+|0110\rangle+|1001\rangle+|1100\rangle)+\gamma_{0-}(|0101\rangle+|1010\rangle)+|1111\rangle$ &$S_0$ &$T_0$ & $H_{0}$\\

\hline
\end{tabular}
\end{footnotesize}
\end{table}
The states with energies $\pm 2(\sqrt{\lambda'}\pm 1)$, that is $|\Psi_{1}\rangle$, $|\Psi_{5}\rangle$, $|\Psi_{10}\rangle$, $|\Psi_{14}\rangle$,  are 
the superposition of two $W$ states  or a local transformation of a $W$ state. As a consequence, $\mathrm{HDet}_{4}$ is zero but not $S$ and $T$.


The states   $|\Psi_{6}\rangle$ and $|\Psi_{9}\rangle$ have  
$\mathrm{HDet}_{4} = 0$, $S \neq 0$ and $T \neq 0$. 
These states entangle  maximally  two spins in one direction with the other spins  in the opposite direction, as explained in section~\ref{sec:Ising}.

Finally, there are four states with $\mathrm{HDet}_{4}$ different from zero. $|\Psi_{0}\rangle$ and $|\Psi_{15}\rangle$ gives the same expression for $\mathrm{HDet}_{4}$ and similarly $|\Psi_{2}\rangle$ and $|\Psi_{13}\rangle$. Figure \ref{Fig:IsThHdet} shows the two $\mathrm{HDet}_{4}$: the corresponding to the second excited state is seven orders of magnitude greater than the corresponding to the ground state, as explained with more detail in section \ref{sec:Ising}.

\section{XXZ model eigenstates\label{app:XXZ}}

The XXZ spin chain with 4 sites  can be solved  analytically as the Ising model. 
Table \ref{Tab:XXZ}  collects 
 the eigenvalues, eigenstates and the $S, T, \mathrm{HDet}_{4}$ invariants. 
 In this case, $|0\rangle$ and $|1\rangle$ denotes  the eigenstates of $\sigma^z$.

For $\Delta<-1$, the ground state is doubly degenerate with an energy 
$4\Delta$; it describes a ferromagnetic phase where all spins are aligned.
For $\Delta>-1$ its energy is $-2\left(\Delta+\sqrt{8+\Delta^2}\right)$ and the ground state is a resonating valence bound, as explained in the main text. At $\Delta=-1^{+}$ is full degenerate and it is a superposition of all spin configurations.

\begin{table}[h!]
\begin{footnotesize}
\caption{\footnotesize$S$, $T$ and $\mathrm{HDet}_{4}$ for states of XXZ model as a function of anisotropy parameter $\Delta$. All states lead to zero $\mathrm{HDet}_{4}$ and only four states have $S$ and $T$ invariants different from zero. Expressions of $S_{\pm}$ and $T_{\pm}$ are written in \eqref{eq:ST_XXZ}.
\label{Tab:XXZstates}
}
\begin{tabular}{C{3cm} C{6cm} C{1.5cm}C{1.5cm}c}
\hline
\textbf{Energy} & \textbf{State} & $\mathbf{S}$  & $\mathbf{T}$  & $\mathbf{HDet}_{\mathbf{4}}$ \\
\hline
-4 & $|0111\rangle -|1011\rangle +|1101\rangle -|1110\rangle$& 0 & 0 & 0 \\
-4 & $|0001\rangle -|0010\rangle +|0100\rangle -|1000\rangle$ & 0 & 0 & 0 \\
4 & $|0111\rangle +|1011\rangle +|1101\rangle +|1110\rangle$ & 0 & 0 & 0 \\
4 & $|0001\rangle +|0010\rangle +|0100\rangle +|1000\rangle$ & 0 & 0 & 0 \\
0 & $|0111\rangle -|1101\rangle$ & 0 & 0 & 0 \\
0 & $|1011\rangle -|1110\rangle$ & 0 & 0 & 0 \\
0 & $|1001\rangle -|1100\rangle$ & 0 & 0 & 0 \\
0 & $|0001\rangle - |0100\rangle$ & 0 & 0 & 0 \\
0 & $|0110\rangle -|1100\rangle$ & 0 & 0 & 0 \\
0 & $|0010\rangle -|1000\rangle$ & 0 & 0 & 0 \\
0 & $|0011\rangle -|1100\rangle$ & $1/(2^6 3)$ & --$1/(2^9 3^3)$ & 0 \\
$-4\Delta$ & $|0101\rangle -|1010\rangle$ & $1/(2^6 3)$ & --$1/(2^9 3^3)$ & 0 \\
$4\Delta$ & $|0000\rangle$ & 0 & 0 & 0 \\
$4\Delta$ & $|1111\rangle$ & 0 & 0 & 0 \\
$-2\left(\Delta-\sqrt{8+\Delta^2}\right)$ & $|0011\rangle +|0110\rangle +|1100\rangle +|1001\rangle-\frac{1}{2}\left(\Delta-\sqrt{8+\Delta^2}\right)\left(|0101\rangle+|1010\rangle\right)$ & $S_{+}$ &$T_{+}$ & 0\\
$-2\left(\Delta+\sqrt{8+\Delta^2}\right)$ & $|0011\rangle +|0110\rangle +|1100\rangle +|1001\rangle-\frac{1}{2}\left(\Delta+\sqrt{8+\Delta^2}\right)\left(|0101\rangle+|1010\rangle\right)$& $S_{-}$ & $T_{-}$ & 0 \\
\hline
\end{tabular}
\end{footnotesize}
\end{table}

For all states obtained after the Hamiltonian diagonalization $\mathrm{HDet}_{4}$ is zero, and only for four states $S$ and $T$ invariants are non zero. Two of these states correspond with the two configurations that maximally entangled two spins up with two spins down:
\begin{eqnarray}
\frac{1}{\sqrt{2}}\left(|0011\rangle-|1100\rangle\right)&=&\frac{1}{\sqrt{2}}\left(|\upuparrows\rangle_{12}|\downdownarrows\rangle_{34}-|\downdownarrows\rangle_{12}|\upuparrows\rangle_{34}\right), \\
\frac{1}{\sqrt{2}}\left(|0101\rangle-|1010\rangle\right)&=&\frac{1}{\sqrt{2}}\left(|\upuparrows\rangle_{13}|\downdownarrows\rangle_{24}-|\downdownarrows\rangle_{13}|\upuparrows\rangle_{24}\right),
\end{eqnarray}
where $|\upuparrows\rangle=|00\rangle$ and $|\downdownarrows\rangle=|11\rangle$.

The other two states are the ones with energies $-2\left(\Delta+\sqrt{8+\Delta^2}\right)$ and $-2\left(\Delta-\sqrt{8+\Delta^2}\right)$, \eqref{eq:XXZ_gs} and \eqref{eq:XXZ_e15}, and correspond respectively with the ground state and $15^{\mathrm{th}}$ excited state for $-1<\Delta<1$.

States that can be factorized into two subsystems have energy zero and $S$ and $T$ zero. These states are
\begin{eqnarray}
\frac{1}{\sqrt{2}}\left(|0111\rangle-|1101\rangle\right)&=&|\Psi^-\rangle_{13}|11\rangle_{24}\nonumber\\
\frac{1}{\sqrt{2}}\left(|1011\rangle-|1110\rangle\right)&=&|11\rangle_{13}|\Psi^-\rangle_{24}\nonumber\\
\frac{1}{\sqrt{2}}\left(|1001\rangle-|1100\rangle\right)&=&|10\rangle_{13}|\Psi^-\rangle_{24}\nonumber\\
\frac{1}{\sqrt{2}}\left(|0110\rangle-|1100\rangle\right)&=&|\Psi^-\rangle_{13}|10\rangle_{24}\nonumber\\
\frac{1}{\sqrt{2}}\left(|0001\rangle-|0100\rangle\right)&=&|00\rangle_{13}|\Psi^-\rangle_{24}\nonumber\\
\frac{1}{\sqrt{2}}\left(|0010\rangle-|1000\rangle\right)&=&|\Psi^-\rangle_{13}|00\rangle_{24}
\end{eqnarray}

Finally, states with energy $\pm 4$ are $W$-type and, consequently, have $S$ and $T$ zero. States with energy 4 have the typical form of a $W$ state and states with energy $-4$ correspond to the local operation $\sigma_z^1\sigma_z^3|W\rangle$, where $\sigma_z^{i}$ is the Pauli matrix operation over $i$-th qubit.

\subsection{Degeneracy}

In the case of degeneracy, a linear combination of eigenstates with the same energy is also an eigenstate. In that case, the values for $\mathrm{HDet}_{4}$, $S$ and $T$ invariants can be altered.

As example, let us analyze the case of Heisenberg model, i.e. XXZ model with $\Delta=1$.

As it is shown in Table \ref{Tab:XXZ_Spin}, there are four different energies in this particular case. The ground state is not degenerate, so the values of $\mathrm{HDet}_{4}$, $S$ and $T$ invariants remain the same as computed in the text: $\mathrm{HDet}_{4}=0$, $S=S_{-}$ and $T=T_{-}$.

The state with energy $E=-4$ has degeneracy 3. Any state with the form
\begin{eqnarray}
|\Psi(E=-4)\rangle &=& \frac{1}{\mathcal{N}}\left(a(|0111\rangle -|1011\rangle +|1101\rangle -|1110\rangle) + b(|0101\rangle -|1010\rangle) \right. \nonumber\\
&&\left. + c(|0001\rangle -|0010\rangle +|0100\rangle -|1000\rangle)\right)
\end{eqnarray}
is also an eigenstate. This state has $\mathrm{HDet}_{4}=0$ due to an exact cancellation between the two invariants:
\begin{eqnarray}
S(E=-4)&=&\frac{(b^2 - 4 ac)^4}{192 (2 a^2 + b^2 + 2 c^2)^4}, \nonumber\\
T(E=-4)&=&-\frac{(b^2 - 4 ac)^6}{13824 (2 a^2 + b^2 + 2 c^2)^6}.
\end{eqnarray}

The state with energy $E=4$ has degeneracy 5. Then, any state with the form
\begin{eqnarray}
|\Psi(E=4)\rangle &=& \frac{1}{\mathcal{N}}\left(a(|0111\rangle +|1011\rangle +|1101\rangle +|1110\rangle) + b|0000\rangle+ \right. \nonumber\\
&& \left. c(|0001\rangle +|0010\rangle +|0100\rangle +|1000\rangle) + d|1111\rangle  + \right. \nonumber\\
&&\left. e(|0011\rangle +|0110\rangle +|1100\rangle +|1001\rangle +|0101\rangle+|1010\rangle)\right)
\end{eqnarray}
is also an eigenstate. In this case, $\mathrm{HDet}_{4}$ could be different from zero. We do not include the expressions of the invariants as they are cumbersome and not very illustrative.

Finally, the state with energy $E=0$ has degeneracy 7. In this case, $\mathrm{HDet}_{4}=0$ again for the cancellation between $S$ and $T$ invariants.

\end{appendix}

\end{document}